\documentclass[twocolumn,showpacs,preprintnumbers,
               superscriptaddress]{revtex4}
\usepackage{graphicx}

\newcommand{\sign}{\mbox{sign}}

\begin{document}

\preprint{UTCCP-P-132, HUPD-0210, YITP-02-66}
\preprint{hep-lat/0211003}

\title{Charmonium at finite temperature in quenched lattice QCD}

\author{Takashi~Umeda}
\affiliation{Center for Computational Physics, University of Tsukuba,
             Tsukuba 305-8577, Japan}

\author{Kouji~Nomura}
\affiliation{Department of Physics, Hiroshima University,
             Higashi-hiroshima 739-8526, Japan}

\author{Hideo~Matsufuru}
\affiliation{Yukawa Institute for Theoretical Physics,
             Kyoto University, Kyoto 606-8502, Japan}

\date{2 November 2002}


\begin{abstract}
We study charmonium correlators in pseudoscalar and
vector channels at finite temperature
using lattice QCD simulation in the quenched approximation.
Anisotropic lattices are used in order to have sufficient numbers of degrees
of freedom in the Euclidean temporal direction.
We focus on the low energy structure of the spectral function,
corresponding to the ground state in the hadron phase,
by applying the smearing technique to enhance the 
contribution to the correlator from this region.
We employ two analysis procedures:
the maximum entropy method (MEM) for the extraction of the spectral
function without assuming a specific form, to estimate the shape
of the spectral function,
and the standard $\chi^2$ fit analysis using typical forms
in accordance with  the result of MEM, for a more quantitative evaluation.
To verify the applicability of the procedures,
we first analyze the smeared correlators as well as the point
correlators at zero temperature.
We find that by shortening the $t$-interval used for the analysis 
(a situation inevitable at $T>0$) the reliability of MEM for
point correlators is lost, while it subsists for smeared correlators.
Then the smeared correlators at $T\simeq 0.9 T_c$ and $1.1 T_c$
are analyzed.
At $T\simeq 0.9 T_c$, the spectral function exhibits a strong peak,
well approximated by a delta function corresponding to the ground
state with almost the same mass as at $T=0$.
At $T\simeq 1.1 T_c$, we find that the strong peak structure still
persists at almost the same place as below $T_c$, but with a finite
width of a few hundred MeV.
This result indicates that the correlators possess a nontrivial
structure even in the deconfined phase.
\end{abstract}

\maketitle

\section{Introduction}
 \label{sec:Introduction}

The charmonium systems have been payed much attention as a signal
of the QCD phase transition.
Due to a change of the interquark potential by thermal effects,
the $J/\psi$ mass is expected to decrease when approaching the
phase transition \cite{Has86}.
Above $T_c$, the screening of the interquark potential may dissolve
the charmonium states, and the resulting $J/\psi$ suppression
has been regarded as one of the most important signals for detecting
the formation of the plasma state \cite{MS86,Jpsi:exp,Jpsi:th}.

In principle, lattice QCD simulations can provide information on
such excitation modes based on a nonperturbative and model independent
framework, since
the correlation functions in the Euclidean temporal direction
measured on a lattice
are related to the real time retarded and advanced Green functions
by analytic continuation from a single spectral function
\cite{AGDF59,HNS93}.
In practice, however, the extraction of reliable information from
numerical data for correlators becomes increasingly
difficult as the temperature increases.
There are two restrictions on the correlators at finite temperature:
the numbers of available degrees of freedom,
and the physical extent in the temporal direction.
The former can be improved by the use of anisotropic lattices,
on which the temporal lattice spacings are finer than the
spatial ones \cite{HNS93,TARO01,Ume01}.
The latter restriction is, however, physically inevitable and 
may spoil standard procedures applicable at zero temperature.

Recently, there has been technical progress in the direct extraction
of the spectral function from lattice data of correlators
\cite{NAH99,SpF_early}.
The key role is played by the maximum entropy method (MEM),
combined with the singular value decomposition for constructing
the functional bases in the space of spectral functions.
At zero temperature this technique has reproduced 
spectra consistent with the standard fit analyses
\cite{NAH99,ODS00,WK00,CPPACS01a,Fie02}.
Application to systems at finite temperature is, however,
not straightforward because of the two restrictions mentioned
above \cite{AMR02}.
In particular, shortness of temporal extent, $1/T$, makes it
difficult to extract the correct low energy structure of the
spectral function.
Therefore, one at least needs to verify at $T=0$ that the method 
produces results which are stable against variation of the $t$-interval.
There have already been a number of applications of MEM
for finite temperature \cite{Bielefeld02a,Bielefeld02b,Bielefeld02c,AHN02},
but as long as such tests are not systematically performed these 
results may contain uncontrolled uncertainties.
In fact, we will show in this paper that MEM applied to
point-point correlators at $T=0$ with shortened $t$-interval
fails to produce the correct spectral functions.
We therefore apply the smearing technique
to enhance the contribution of the low energy part to the correlators
\cite{ODS00,TARO01,Ume01}.
For the smeared correlators at $T=0$, MEM works, at least qualitatively,
also with a shorter $t$-interval.

In this paper, charmonium correlators are investigated using lattice
QCD simulations in the quenched approximation.
We focus on properties of the ground state, more generally the low energy
structure of the correlator, rather than the whole 
spectral function.
This paper mainly deals with the following two subjects:

(1) A technical investigation of procedures for extraction of
reliable information from the numerical data of correlators in the
temporal direction.
In addition to MEM, we apply the standard $\chi^2$ fit analysis
assuming a peak structure for the spectral function \cite{ISM02}.
The latter approach complements the former, since MEM with small
numbers of data points appears not work beyond the level of qualitative
estimate for the shape of spectral function.

(2) A study of temperature effects on the correlators near to
the deconfining phase transition.
Below $T_c$, the question is whether the charmonium mass shifts,
as expected from the potential model approach \cite{Has86}.
Above $T_c$, previous studies have indicated that nontrivial
structures may survive \cite{TARO01,Ume01}.
The possibility of the existence of collective modes in the plasma phase
is therefore carefully examined.

We use quenched anisotropic lattices with the renormalized anisotropy
$\xi=a_\sigma/a_\tau = 4$ and the spatial lattice cutoff
$a_\sigma^{-1}\simeq 2$ GeV.
The phase transition occurs near  $N_t=28$,
and hence sufficiently many points are available
in the vicinity of $T_c$.
At $T\simeq 0$ we try to determine the conditions 
for a reliable extraction of the spectral function.
We find that for the point correlators, reliability of MEM with 
less than 24 $t$-points (meaning a physical $t$-range
of less than $O(0.5 \mbox{fm}) \sim 0.7 T_c$) is not guaranteed
with the present level of statistics.
Therefore only the smeared correlators are analyzed at
finite temperature, $T\simeq 0.9 T_c$ and 1.1 $T_c$.

Here we comment on the smearing technique.
Although the correlator of spatially extended operators has
no direct relation to the physically observable dilepton
production rate, the properties of collective excitations such
as mass and width are unchanged and can be probed more efficiently.
To the extent that these excitations drive the dilepton rate we
can at least say something about their position and shape, if not 
about their strength.
A disadvantage is the possibility of detecting a fake peak,
which may be produced by the smearing even in the case of free quarks
\cite{Bielefeld02a}.
To distinguish such an artifact from a genuine physical peak
we use besides the smearing function based on the wave
function obtained at $T=0$ also a narrower one of essentially half-width.
We speak thereby of ``smeared'' and of  ``half-smeared'' correlators
(because the energy region enhanced by this narrower smearing function 
is wider, while the high frequency part of $O(1/a_\tau)$
is still sufficiently suppressed).
Comparing the results of these two types of smeared correlators,
one can argue whether the observed peak structure is an artifact
of  free quarks or physical indication of a collective mode.

This paper is organized as follows.
In Sec.~\ref{sec:spectral_func}, we recall some  fundamental properties
of the spectral function.
For later uses in the $\chi^2$ fit analysis, presumable forms
of the spectral function for a collective mode are discussed.
Section~\ref{sec:action} reviews anisotropic lattice actions
and describes the parameters used in this paper.
We also show the spectral function of the correlator of a
free quark-antiquark pair, for comparison with the 
correlators from
the Monte Carlo simulation.
In Sec.~\ref{sec:procedure} our analysis procedures are explained.
Section~\ref{sec:simulation} describes the setup of Monte Carlo
simulation.
The following three sections show the result of the numerical simulation.
In Section~\ref{sec:result_zt} we analyze the correlators at
zero temperature, as a preparation for the more involved  situations
at finite temperatures.
Sections~\ref{sec:result_ft1} and \ref{sec:result_ft2} show
the results at $T\simeq 0.9 T_c$ and $1.1 T_c$, respectively.
In each case, we first estimate the presumable form of the spectral
function with MEM, and then apply the $\chi^2$ fit analysis for
a more quantitative evaluation of its structure.
Section~\ref{sec:conclusion} is devoted to our conclusions
on the technical and physical implications of the results.
Our preliminary results have been reported in Ref.~\cite{NMUM02}.

\section{Spectral function}
\label{sec:spectral_func}

In this section $\beta$ represents the inverse temperature and
should not be confused with the coupling of gauge field which
appears in later sections.
For a finite lattice of temporal extent $N_\tau$ we have
 $\beta = 1/T = N_t a_\tau$.
The Euclidean time is represented by $\tau$ in this section
to make clear the distinction 
 between the Euclidean and Minkowski formulations.
In later sections we use also $t$ as the Euclidean time,
because no confusion is expected for the lattice QCD
formulated in the Euclidean space.

We consider correlators of the form:
\begin{equation}
 C(\tau) = \sum_{\vec{x}} \langle
       O(\vec{x},\tau) O^{\dag}(0,0)   \rangle ,
 \label{eq:correlator}
\end{equation}
where the operator $O(\vec{x},\tau)$ is a quark bilinear,
\begin{equation}
 O(\vec{x},\tau) = \sum_{\vec{y}} \phi(\vec{y})
   \bar{q}(\vec{x}+\vec{y},\tau) \Gamma q(\vec{x},\tau) .
 \label{eq:operator}
\end{equation}
4$\times$4 matrix $\Gamma$ specifies the quantum number, such
as, for example, $\Gamma=\gamma_5$ for pseudoscalar and
$\gamma_i$ ($i=1,2,3$) for vector channels.
$\phi(\vec{y})$ is a smearing function defining the  spatial extension of
the bilinear operator.
Since the smearing in general violates the gauge invariance of
the correlator, one needs to fix the gauge or employ a gauge invariant
form of the smearing function.
We fix the configurations to the Coulomb gauge, since it is
widely used and allows easy implementation of the smearing function.
Source and sink
operators are always identical in this work.

{\it Point} correlators correspond to a ultra local ``smearing function'' 
$\phi(\vec{y})=\delta_{\vec{y},0}$ in Eq.~(\ref{eq:operator}).
For the {\it smeared} correlators we use smearing function of the form
\begin{equation}
 \phi(\vec{y}) = \exp(-a |\vec{y}|^p),
 \label{eq:wavefunction}
\end{equation}
where the parameters $a$ and $p$ are determined by fitting
the wave function measured in the numerical simulation
to the above form.
The smearing technique described here was already applied to
the problems of the charmonium correlators at $T>0$ \cite{Ume01},
as well as to the light meson correlators \cite{TARO01}.

For a while let us consider the continuum Euclidean space.
The Matsubara Green function is represented in terms of the
spectral function as
\begin{equation}
 G(\nu_n , \vec{p})
  = \int_{-\infty}^{\infty} \frac{d\omega}{2\pi}
    \frac{\rho (\omega , \vec{p})}{ i \nu_n - \omega + i\epsilon} .
\end{equation}
The spectral function $\rho(\omega ,\vec{p})$ is represented as
\begin{eqnarray}
\rho(\omega, \vec{p}) &=& \frac{1}{2 Z(\beta)} \sum_{n,m}
      ( e^{-\beta E_n} - e^{-\beta E_m} )
      |\langle n | O(0) | m \rangle |^2
  \nonumber\\
 & & \hspace{0.0cm} \times (2\pi)^4 \delta( \omega - E_m + E_n )
                       \delta( \vec{p} - \vec{P}_m + \vec{P}_n ),
 \label{eq:specfunc}
\end{eqnarray}
which is an odd real function due to the bosonic nature of the
present correlator and positive for $\omega >0$.

The retarded and advanced Green functions for real time
are represented as
\begin{eqnarray}
G^R(\nu,\vec{p})
 &=& \int_{-\infty}^{\infty}
   \frac{d\omega}{2\pi}
   \frac{\rho (\omega,\vec{p})}{\nu - \omega + i\epsilon}
 \\
G^A(\nu,\vec{p})
 &=& \int_{-\infty}^{\infty}
   \frac{d\omega}{2\pi}
   \frac{\rho (\omega,\vec{p})}{\nu - \omega - i\epsilon}
\end{eqnarray}
with the same spectral function, $\rho(\omega,\vec{p})$
\cite{AGDF59}.
The spectral function is then represented as
\begin{equation}
 \rho(\nu, \vec{p}) = - 2 \; \mbox{Im} G^R(\nu,\vec{p})
          =   2 \; \mbox{Im} G^A(\nu,\vec{p}) .
\end{equation}
Strictly speaking, the above integrations do not always converge,
and one needs appropriate subtraction terms.
On the other hand, the high frequency part of the spectral function is
not practically significant in the numerical analysis,
because of the strong suppression by $K(\tau,\omega)$,
and of the existence of the lattice cutoff.
We therefore do not consider these subtractions in this paper.
We note that the smearing of the operator in Eq.~(\ref{eq:operator})
is performed only in the spatial directions, and therefore  the analytic
continuation of the Matsubara Green function to the real time Green
functions is independent of this operation (but, of course, the
spectral function depends on the operators, hence on smearing).

In the following  we are only concerned with the correlators projected on  the
zero momentum states and discard the argument $\vec{p}$.
Since the spectral function $\rho (\omega)$ is an odd function in
$\omega$, the correlator (\ref{eq:correlator}) is represented as
\begin{equation}
C(\tau) =  \int_{0}^{\infty}
   \frac{d\omega}{2\pi}
   K(\tau, \omega) \rho(\omega) .
\label{eq:spec_repr}
\end{equation}
where
\begin{equation}
K(\tau, \omega)
 = \frac{ e^{-\omega \tau} + e^{- \beta \omega + \omega \tau} }
                 {1 - e^{-\beta \omega}} .
\label{eq:kernel}
\end{equation}
In the numerical simulation of lattice QCD one measures
the left hand side of Eq.~(\ref{eq:spec_repr}).
The question  is how to  solve the inverse problem to obtain
$\rho (\omega)$ with limited data for   $C(\tau)$.
The procedures to extract information about  $\rho (\omega)$
are described in Sec.~\ref{sec:procedure}.

Now we shall present several {\it ans\"atze} for the spectral function
which will be used for later analysis of the lattice data, and discuss 
their physical implications.
A simple representation of a collective mode
is the following form of the retarded Green function:
\begin{equation}
G^R(\nu)
 = \frac{R(\nu)}{ \nu - \mu + i\gamma(\nu)/2},
\end{equation}
where $\mu$ and $\gamma(\nu)$ express the
dispersion and the width of the mode.
The corresponding spectral function is the well-known Breit-Wigner
form.
We shall neglect the frequency dependence of the width $\gamma$
for simplicity.
Taking the oddness of the bosonic spectral function into account,
$\rho(\omega)$ reads
\begin{eqnarray}
 \rho(\omega) =
    \frac{\gamma R(\omega)}{(\omega - m )^2 + \gamma^2/4 }
  - \frac{\gamma R(\omega)}{(\omega + m )^2 + \gamma^2/4 }.
\label{eq:spc_BW}
\end{eqnarray}
If the quantum numbers of the operators used to represent
the collective mode are the correct ones, the physical properties
associated with the mode, characterized by $\mu$ and $\gamma$,
are independent of the particular operator.
Therefore the smeared operators can be used to observe these
quantities.
However, the smearing does change the overlap of the operator
with the state, $R(\omega)$.
Therefore the effects depending on  $R(\omega)$ and those coming
only from the peak structure (represented
with $\mu$ and $\gamma$) should be distinguished.
If the change of $R(\omega)$ is sufficiently gentle over the region
of the observed peak, the $\omega$ dependence in $R(\omega)$ is
negligible.

Instead of Eq.~(\ref{eq:spc_BW}), at zero temperature
one often uses the relativistic Breit-Wigner type form:
\begin{equation}
 \rho(\omega) = \frac{\sign (\omega) \omega^2 m \gamma (4R/m)}
                     {(\omega^2 - m^2 )^2 + m^2 \gamma^2 }.
\label{eq:spc_RBW}
\end{equation}
This form shows a similar behavior to Eq.~(\ref{eq:spc_BW})
around the peak position.
However, the behavior far from the peak is different.
In particular (\ref{eq:spc_BW}) and (\ref{eq:spc_RBW})
give different contribution to the integral
(\ref{eq:spec_repr}) near vanishing $\omega$:
In the vicinity of $\omega = 0$,  Eq.~(\ref{eq:spc_BW}) behaves
linearly in $\omega$,
while Eq.~(\ref{eq:spc_RBW}) behaves linearly in $\omega^2$.
Since $K(\tau, \omega)$ behaves as $(2-\beta\omega)/\beta\omega$,
Eq.~(\ref{eq:spc_BW}) gives a $t$-independent contribution to
the integral (\ref{eq:spec_repr}) \cite{AMR02}.
We note that there is no indication of linear behavior in
$\omega$ in the small $\omega$ region of the spectral functions
obtained from our numerical data with MEM,
as shown in Sec.~\ref{sec:result_ft1} and \ref{sec:result_ft2}.
Therefore, in the $\chi^2$ fit analysis, we use the form of
Eq.~(\ref{eq:spc_RBW}) as {\it ansatz} for the spectral function
to which the lattice data are fitted.
While the use of Eq.~(\ref{eq:spc_RBW}) is valid at zero temperature,
it should only be taken as  a representative form of peak structure
at $T>0$,
and hence the parameters $m$ and $\gamma$ do not have a strict sense
but represent only  a convenient parameterization of the peak.

In the limit of $\gamma \rightarrow 0$, both Eqs.~(\ref{eq:spc_BW})
and (\ref{eq:spc_RBW}) tend towards a delta function form,
\begin{equation}
 \rho(\omega) = 2\pi R \left[   \delta(\omega-m)
                              - \delta(\omega+m) \right].
\label{eq:spc_delta}
\end{equation}
This form corresponds to the standard exponential fit
of correlators used in the hadron spectroscopy at zero temperature.

\section{Anisotropic lattice QCD}
 \label{sec:action}

\subsection{Anisotropic lattice actions}

For analysis of the correlators in the temporal direction
at $T>0$ the fine temporal resolution is very important
to extract meaningful information \cite{TARO01, Ume01, ISM02}.
Anisotropic lattices have become a powerful tool to achieve
sufficiently high temporal cutoff while keeping total computational
size modest.
In the following, we summarize the anisotropic lattice actions
employed in this work.
The parameters we use are described in the next subsection.

For the gauge field, we adopt the standard Wilson plaquette
action \cite{Kar82},
\begin{eqnarray}
 S_G &=& \beta \gamma_G \sum_{x} \sum_{i=1}^{3}
        \left[ 1 - \frac{1}{3} \mbox{ReTr}U_{i4}(x) \right]
 \nonumber \\
     & & + \frac{\beta}{\gamma_G} \sum_{x} \sum_{i<j=1}^{3}
        \left[ 1 - \frac{1}{3} \mbox{ReTr}U_{ij}(x) \right] ,
\end{eqnarray}
where $U_{\mu\nu}$ is a product of link variables
along a plaquette in the $\mu$-$\nu$ plane.
Here the parameters $\beta$ and $\gamma_G$ are bare coupling and bare
anisotropy, respectively.

For the quark field we use the $O(a)$ improved Wilson-type
action \cite{Ume01,Aniso01a,Aniso01b}:
\begin{equation}
 S_q = \sum_{x,y} \bar{q}(x) K(x,y) q(y), 
\end{equation}
\begin{eqnarray}
 K(x,y) &=& \delta_{xy}
      \nonumber \\
 & &  \hspace{-1.4cm}
    - \kappa_{\tau} \left[ (1\!-\!\gamma_4)U_4(x) \delta_{x+\hat{4},y}
        + (1\!+\!\gamma_4)U_4(x-\hat{4}) \delta_{x-\hat{4},y} \right] 
      \nonumber \\
 & &  \hspace{-1.4cm}
    - \kappa_{\sigma} \sum_{i=1}^3
     \left[ (r\!-\!\gamma_i)U_i(x) \delta_{x+\hat{i},y}
      + (r\!+\!\gamma_i)U_i(x\!-\!\hat{i}) \delta_{x-\hat{i},y} \right] 
       \nonumber \\
 & &
   - \kappa_{\sigma} c_E \sum_{i=1}^3 \sigma_{4i} F_{4i}(x) \delta_{x,y}
     \nonumber \\
 & &
  - \kappa_{\sigma} c_B \sum_{i>j=1}^3 \sigma_{ij} F_{ij}(x) \delta_{x,y}.
\label{eq:qaction}
\end{eqnarray}
where $\kappa_{\sigma}$ and $\kappa_{\tau}$ are the spatial
and temporal hopping parameters, respectively, $r$ is the spatial Wilson parameter,
and $c_E$,  $c_B$ are the clover coefficients for the $O(a)$
improvement.
We set $r=1/\xi$ and $c_E$, $c_B$ to the tadpole-improved tree-level
values, and vary the two parameters $\kappa_\sigma$ and
$\kappa_\tau$ to change the quark mass and the fermionic anisotropy.
The tadpole improvement \cite{LM93} is performed by rescaling the
link variables as $U_i(x) \rightarrow U_i(x)/u_{\sigma}$ and
$U_4(x) \rightarrow U_4(x)/u_{\tau}$, respectively,
with the corresponding mean-field values of the spatial and temporal link variables,
$u_{\sigma}$ and $u_{\tau}$.
Then $c_E$ and $c_B$ with the choice $r=1/\xi$ read as
\begin{equation}
 c_E= 1/u_{\sigma} u_{\tau}^2, \hspace{0.5cm}
 c_B = 1/u_{\sigma}^3 .
\label{eq:cecb}
\end{equation}
Instead of $(\kappa_\sigma, \kappa_\tau)$, it is convenient to
use the parameters
\begin{eqnarray}
 \gamma_F &\equiv& u_\tau \kappa_{\tau} / u_\sigma \kappa_{\sigma},\\
 \frac{1}{\kappa} &\equiv& \frac{1}{u_\sigma \kappa_{\sigma}}
     - 2(\gamma_F + 3r - 4)
    = 2( m_{0}\gamma_F + 4),
 \label{eq:kappa}
\end{eqnarray}
where $m_0$ is the bare quark mass in temporal lattice units.
$\kappa$ plays the same role as on isotropic lattices, and controls
the bare quark mass.
$\gamma_F$ is the bare fermionic anisotropy.

The lattice quark field $q(x)$ in the action~(\ref{eq:qaction})
is related to the
dimensionful field $\psi(x)$ as
\begin{equation}
 \psi(x) = a_{\sigma}^{-3/2} Z_q(\beta,\kappa) {\cal K}(\kappa) q(x),
\label{eq:field_dim}
\end{equation}
where we assume $m_Q \ll a_\tau^{-1}$.
${\cal K}(\kappa)$ is so-called KLM normalization factor \cite{KLM},
\begin{eqnarray}
 {\cal K}(\kappa) &=& [ 2 u_\tau \kappa_\tau (1 + m_0) ]^{1/2}
 \nonumber \\
  &=& \left[ \frac{1 + 2\kappa (\gamma_F - 4)}
                  {1 + 2\kappa (\gamma_F + 3r - 4)} \right]^{1/2} .
\label{eq:KLM_factor}
\end{eqnarray}
We consider the tadpole-improved tree-level matching of the fields
in continuum and lattice theories, and hence $Z_q$ is set to unity.

\subsection{Parameters}
 \label{subsec:calibration}

We here describe the parameters used in this paper.
First, we discuss the calibration of gauge and quark fields.
In an interacting theory, the renormalized anisotropy
$\xi = a_\sigma/a_\tau$ generally differs from the
bare anisotropy parameters $\gamma_G$ and $\gamma_F$.
Therefore one needs to tune the latter such  that
\begin{eqnarray}
  \xi_G(\gamma_G,\gamma_F) = \xi_F(\gamma_G,\gamma_F) = \xi
\end{eqnarray}
holds, where $\xi_G$ and $\xi_F$ are the observed anisotropies
defined through the gauge and fermionic observables, respectively.
In quenched simulations, one can calibrate the gauge and fermion
fields separately, first the former, and then the latter.

For the calibration of the gauge field,
we refer to an elaborated work by Klassen \cite{Kla98} which
uses Wilson loops in the spatial-spatial and spatial-temporal
planes and is performed at 1\%  accuracy level.
Making use of his relation of $\gamma_G$ to $\beta$ and $\xi$,
we choose $\beta=6.10$ and $\gamma_G=3.2108$ corresponding
to the renormalized anisotropy $\xi=4$.
These values were also adopted in Ref.~\cite{Aniso01b},
and correspond to the spatial lattice cutoff
$a_{\sigma}^{-1}=2.030(13)$ GeV set by the hadronic radius $r_0$
\cite{Som94}.

For the quark field, we use the result of Ref.~\cite{Aniso01b}
in which the calibration was performed on the same lattice as this
work using the meson dispersion relation in a quark mass region
including the charm quark mass.
Accordingly, we adopt $\kappa=0.112$ and $\gamma_F=4.00$, namely
$m_q = 0.121$, as the values corresponding to the charm quark mass.
(In Ref.~\cite{Aniso02a} it was argued that the value of $\gamma_F$ tuned
for the massless quark is also applicable to the charm quark mass.
For historical reason, however, we use the result of the mass
dependent tuning in Ref.~\cite{Aniso01b}.
The difference is actually small and does not cause any significant
effect.)

As smearing function we use the result for the wave function in the
Coulomb gauge at $T=0$.
For the vector channel the fit to Eq.~(\ref{eq:wavefunction})
yields $a = 0.2275(9)$ and $p=1.258(5)$.
These values are used for both the pseudoscalar and vector
correlators.

In addition, we also use a narrower smearing function
with $a=0.45$ and the same $p$.
The correlators smeared with this function are called 
{\it half-smeared} correlators, since the smearing function has about
twice the slope as the main one and hence a smaller  smearing effect
is expected: 
the low energy part of the correlator is only partially enhanced,
while the high energy part of $O(1/a_\tau)$ is sufficiently
reduced.
This intermediate smearing function plays an important role
at $T>T_c$ in examining whether the observed peak structure of
the spectral function for the smeared correlator
is an artifact of smearing or a genuine physical effect
(Sec.~\ref{sec:result_ft2}).

\subsection{Free quark case}
 \label{subsec:free_quark}

\begin{figure}
\vspace{-0.35cm}
\includegraphics[width=9.0cm]{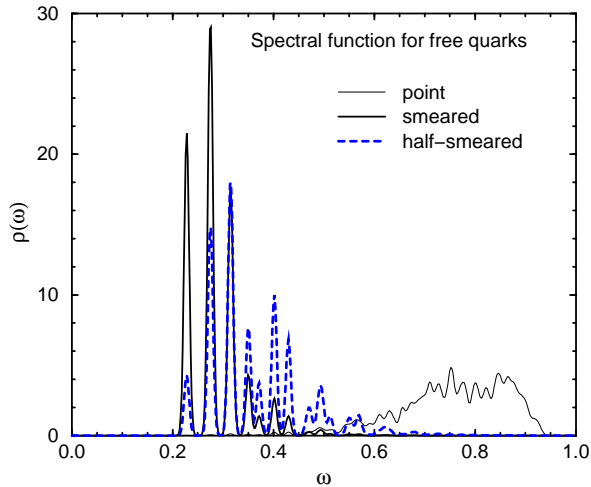}
\vspace{-0.9cm}
\caption{The spectral functions for the pseudoscalar correlators
composed of free quarks at $T=0$.}
\label{fig:spcf}
\end{figure}

Since in the perturbative plasma regime we expect free quarks and gluons
it is important to compare the result of Monte Carlo simulation
with the case of free quarks on the lattice.
The spectral function of the correlator (\ref{eq:correlator})
in the case of free quarks is obtained from the expression
(\ref{eq:specfunc}).
In the case of vanishing total momentum, for positive frequency, 
\begin{eqnarray}
 \rho(\omega)
   &=& \frac{\pi}{Z} \sum_{\vec{p}} [ 1-e^{-N_t 2 E(\vec{p})} ]
                \tilde{\phi}(\vec{p})^2
 \nonumber \\
   & & \times  \sum_{j,k=1,2}
    \left[ \frac{\bar{v}_{j}(-p) \Gamma u_{k}(p)}{2E(p)} \right]^2
            \delta\left( \omega - 2 E(\vec{p}) \right)
\end{eqnarray}
where $u_{j}(p)$ and $v_{j}(p)$ are the quark and antiquark
spinors with $j$-th spin, and $\tilde{\phi}(\vec{p})$ is the
Fourier transform of $\phi(\vec{y})$.
The summation in $\vec{p}$ is taken over all modes on the finite
lattice of the same size as in Monte Carlo simulation, $20^3$.
With the present lattice quark action, the energy of free quark,
$E(\vec{p})$, satisfies the dispersion relation
\cite{Ume01,Aniso01a}
\begin{equation}
 \cosh E(\vec{p}) = 1 + \frac{\vec{\bar{p}}^2 + \left[
           m_0 + \frac{r}{2\gamma_F}\vec{\hat{p}}^2 \right]^2}
          { 2 \left[ 1 + m_0 + \frac{r}{2\gamma_F}\vec{\hat{p}}^2
                                                          \right] },
 \label{eq:freeq_dr}
\end{equation}
where $\bar{p}_i=(1/\gamma_F)\sin p_i$ and $\hat{p}_i=2\sin(p_i/2)$.
We set the bare quark anisotropy $\gamma_F = \xi=4$ and
the bare quark mass $m_0=0.121$ such that the free quark has
the same value as in the Monte Carlo simulation.

Figure~\ref{fig:spcf} shows the spectral functions of
point, smeared, and half-smeared correlators composed of free quarks
in  PS channel.
The integral of these functions are normalized to unity.
Since on the finite lattice the spectral function is a sum of delta
functions, we represent each contribution at  momentum $\vec{p}$
with a Gaussian of width 0.005.
As clearly shown in the figure, the smearing of the operator strongly
enhances the low energy part of the spectral function, as compared to
the point correlator.
In the (fully) smeared correlator, several low momentum states
over the range of $\omega$ about 0.1 dominate 
the correlator.
From this correlator, therefore, an analysis with poor energy resolution
in $\omega$ might produce a peak with full width of order of 0.1
instead of a sum of distinct peaks corresponding to
individual momentum states.
As we will see in Sec.~\ref{subsec:analysis_free_quark}, the
$\chi^2$ fit analysis with the single Breit-Wigner type {\it ansatz}
applied to the smeared correlator
composed of free quarks produces a much narrower peak with the width
of about 0.027 centered at $\omega \simeq 0.275$.
This is of the same order of magnitude as the width extracted  from the
correlators in Monte Carlo simulation at $T\simeq 1.1 T_c$.
Therefore this occasion must be examined  carefully.

It is for this purpose that the half-smeared correlator is introduced.
As shown in Fig.~\ref{fig:spcf}, a slightly higher and wider frequency
region contributes dominantly to the half-smeared correlator
than to the smeared one.
If  the correlator is approximated by a single peak,
the peak will be centered at larger energy and have wider width
as compared to those of the smeared correlator.
This behavior is verified in Sec.~\ref{sec:procedure} by applying
the $\chi^2$ fit analysis to the correlators composed of free quarks.
Therefore, comparing the results for spectral functions for
the smeared and half-smeared correlators can help judging whether the peak
 observed from the simulation data is a genuine
excitation mode or an artifact brought by the smearing.

\section{Analysis procedure}
 \label{sec:procedure}

\subsection{Strategy of analysis}

One of the goals of this work is to investigate techniques to extract
the relevant information of the spectral function from the correlators
with limited numbers of the degrees of freedom.
We treat an inverse problem represented as
\begin{equation}
C(t) = \int_0^{\infty} d\omega K(t,\omega) A(\omega),
\label{eq:problem}
\end{equation}
where $C(t)$ is the given lattice result for the correlator,
and the spectral function $A(\omega)$ is what we need to obtain.
Hereafter we shall denote the spectral function 
reconstructed from the lattice data $A(\omega)$
instead of $\rho(\omega)$ ($\times (2\pi)^{-1}$).
Unless stated otherwise, the variables are in temporal lattice units in the following.
The kernel $K(t,\omega)$ is given as
\begin{equation}
 K(t,\omega)=\frac{ e^{-\omega t}+e^{-\omega(N_t-t)} }
                     { 1 - e^{-N_t\omega} }.
\label{eq:cont_kernel}
\end{equation}
This is the continuum type kernel, and an alternative form of the kernel
which explicitly incorporate the lattice structure was also applied
in the literatures
\cite{NAH99,Bielefeld02a}.
We have seen  no significant difference in applying the two kernels,
therefore we shall not further discuss this point.
In practice, the above integration over the frequency $\omega$ is replaced
by a summation with sufficiently fine discretization $\Delta\omega$
and a cut off at some maximum value $\omega_{max}$.

As already mentioned, one of the main analysis procedures is the
maximum entropy method (MEM) \cite{NAH99}.
Before applying it to finite temperature problems, one should
verify its applicability under the condition which one encounters
at $T>0$.
For this  we test how the result changes when varying
the number of data points used in analysis for the correlators
at $T=0$.
The condition to be satisfied is that the ground state peak --
which we know to exist at $T=0$ -- is
correctly reproduced, at least at a qualitative level.
We stress that without such verification, the result of MEM may
contain uncontrolled artifacts  at finite temperature.
Concerning quantitative questions, we find that with reduced
 number of data points, MEM hardly gives a result
beyond the qualitative level, for example for the width of a peak.

Another technique in our hand is the standard $\chi^2$ fit
assuming a certain  {\it ansatz} for the spectral function.
This method has been applied to a problem of glueballs at
finite temperature \cite{ISM02}.
A disadvantage of this approach is that one needs to assume
a form of the spectral function, which introduces a bias.
For this purpose, the result of MEM can be a good guide.
Once a specific {\it ansatz} is used, the $\chi^2$ fit gives much
more reliable results than MEM for the parameters,
both in statistical as well as systematic sense.

Therefore MEM and the $\chi^2$ fit are complementary to each other
at this stage of computational power,
and in combination provide a more reliable way to analyze
the structure of spectral functions than taken independently.
We thus propose a two-step procedure:
we first apply MEM to the correlators, for a rough estimate of the
shape of spectral function.
Once a presumable form is known, we then examine this form 
using $\chi^2$ fit, and estimate the values of parameters such
as the mass and width of a peak structure.

\subsection{Maximum entropy method}
 \label{subsec:procedure_MEM}

Our MEM analysis basically follows Ref.~\cite{NAH99},
which reviews in detail the maximum entropy method applied to data
of lattice QCD simulation.
Here we briefly summarize just several formulae necessary
for the later description of our analysis.

To obtain $A(\omega)$ from $C(t)$ by solving the inverse problem
Eq.~(\ref{eq:problem}),
the maximum entropy method maximizes a functional
\begin{equation}
 Q(A;\alpha) = \alpha S[A] - L[A].
\label{eq:Q}
\end{equation}
$L[A]$ is the likelihood function,
\begin{equation}
 L[A] = \frac{1}{2}\sum_{t_1,t_2} [ C(t_i) - C_A(t_i) ]
                       V(t_i,t_j)^{-1}  [ C(t_j) - C_A(t_j) ],
\label{eq:L}
\end{equation}
where $C_A(t)$ is the resulting correlator  (\ref{eq:problem})
for the trial $A(\omega)$, and $V(t_i,t_j)$ is the covariance matrix of the data
\begin{equation}
 V(t_i,t_j) = \frac{1}{N(N\!-\!1)} \sum_{k=1}^{N}
    \left[ C_k(t_i) - C(t_i) \right] \left[ C_k(t_j) - C(t_j) \right]
\end{equation}
with $C_k(t)$ the $k$-th sample of the correlator.
The standard $\chi^2$ fit minimizes this functional $L$.
The Shannon-Jaynes entropy $S[A]$ is defined as
\begin{equation}
 S[A] = \int_0^\infty d\omega
   \left[ A(\omega)-m(\omega)-A(\omega)
           \log \left( \frac{A(\omega)}{m(\omega)}\right) \right].
\label{eq:S}
\end{equation}
The function $m(\omega)$ is called the default model, and should
be given as a plausible form of $A(\omega)$.
The parameter $\alpha$ can be integrated out at the last stage of
calculation.
Following Ref.~\cite{NAH99}, we use a form
\begin{equation}
 m(\omega)=m_{DM}\omega^2 .
 \label{eq:default_model}
\end{equation}
In the case of point correlators, a natural choice of $m_{DM}$ is
determined according to the asymptotic behavior of the meson
correlators at large $\omega$ in perturbation theory.
Although such an asymptotic behavior is not observed in practical
simulation because of the finite lattice cutoff,
this form has been successfully applied to problems at zero
temperature \cite{NAH99,CPPACS01a}.
For the smeared correlators, the situation is more subtle, since
the high frequency part of the correlator is suppressed by
the smearing function.
We therefore adopt a practical choice: we normalize the smeared
correlator so as to produce the same overlap with the ground state
as the point correlator.
Then the same normalization is also applied to the correlators at
finite temperatures, and we observe how the result changes with the
change of $m_{DM}$.

In the maximization step of $Q(A;\alpha)$ the singular value
decomposition of the kernel $K(t,\omega)$ is used.
Then the spectral function is represented as a linear combination of
the eigenfunctions of $K(t,\omega)$.
The number of degrees of freedom of $A(\omega)$ is accordingly
reduced to the number of data points of the correlator.
Although the coefficients of the linear combination for $A(\omega)$
can in principle be determined uniquely from the data without
introducing an entropy term, the small eigenvalues of $K(t,\omega)$,
$\theta_i$, lead for the spectral function 
to a singular behavior;
hence truncation at some $i$ is practically required \cite{SpF_early}.
In MEM, the addition of the entropy term stabilizes the problem and
guarantees an unique solution for the coefficients of the eigenfunctions
\cite{NAH99}.
In our analysis, we use only the basis functions for which
$\theta_i > 10^{-12}\times \theta_1$ holds, where $\theta_i$ is
$i$-th largest singular value.
This restriction has no significant effect on the result.

\subsection{$\chi^2$ fit with {\it ansatz} for the spectral function}

The standard $\chi^2$ fit method minimizes the likelihood function
$L$, Eq.~(\ref{eq:L}), with an assumed form for the spectral function.
A most simple form for fit function is the delta function:
\begin{equation}
 A_{pole}(\omega; r,m) = r \delta(\omega-m).
\end{equation}
This form is referred to as {\it pole ansatz} in the following.
At  $T=0$, a sum of several pole terms should describe well the
correlators.
For the correlators below $T_c$, where narrow thermal widths
are expected, the multi-pole form is also convenient to test 
this assumption.

As noted in Sec.~\ref{sec:spectral_func},
to describe a peak structure with finite width we adopt the
relativistic Breit-Wigner form (referred to as {\it BW} form),
\begin{equation}
 A_{BW}(\omega; r, m,\gamma)
    = \frac{\omega^2  r m \gamma}
           {(\omega^2 - m^2 )^2 + m^2 \gamma^2 }.
\end{equation}
This is the same form as Eq.~(\ref{eq:spc_RBW}), with slightly modified
notation for convenience.
This {\it ansatz} neglects the $\omega$ dependence of $r$,
and hence is valid only for the case that the width of the peak,
 $\gamma$,
is sufficiently small compared with the change of the smearing
function in the region of interest.

Combining these two forms, we fit the numerical data of correlators
to the following {\it ans\"atze}:
\begin{description}
\item{\it 2-pole form}:
this form is suitable for description of correlators at $T=0$.
It is also expected to be a good representation of correlators
at $T<T_c$;
\begin{equation}
 A(\omega) =  A_{pole}(\omega; r_0,m_0) + A_{pole}(\omega; r_1,m_1).
\label{eq:fit_2-pole}
\end{equation}

\item{\it 1-BW form}: 
if the contribution from the high frequency part of the spectral function is
negligible, a collective excitation at low energy is expected to be
well represented by a single Breit-Wigner type {\it ansatz}:
\begin{equation}
 A(\omega) =  A_{BW}(\omega; r_0,m_0,\gamma_0).
\label{eq:fit_1-BW}
\end{equation}
This form is also a good representation for the
spectral function of the smeared correlator
composed of free quarks.

\item{\it BW+pole form}:
although the lowest peak is well represented with the Breit-Wigner
type function, there may exist contribution from high frequency
region.
Since the smearing of operator suppress such contribution,
remaining effects of this region may be represented as a single
pole-like term.
This is the reason that we fit the data to the form
\begin{equation}
 A(\omega) =  A_{BW}(\omega; r_0,m_0,\gamma_0)
              + A_{pole}(\omega; r_1,m_1).
\label{eq:fit_BW+pole}
\end{equation}
This is the most general {\it ansatz} for $\chi^2$ fit in this paper.

\end{description}

These {\it ans\"atze} form are the basis for the analysis of the
lowest peak structure, which corresponds to the ground state at $T<T_c$.
The structure in the high frequency region is out of the scope of this 
paper.

\subsection{Analysis of correlators composed of free quarks}
 \label{subsec:analysis_free_quark}

\begin{figure}
\includegraphics[width=8.0cm]{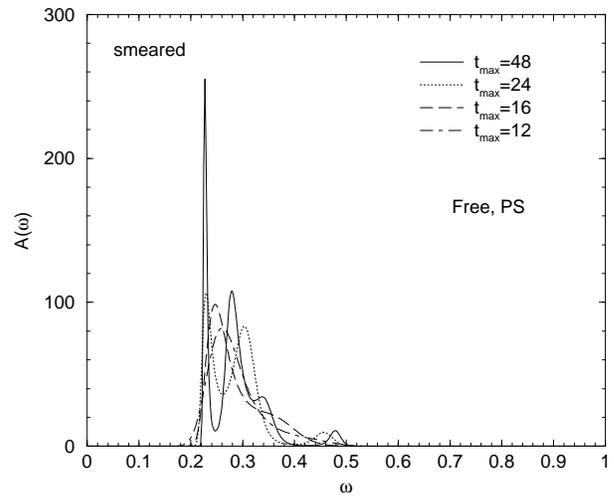}
\caption{The result of MEM analysis for the spectral function
from the smeared PS correlator composed of free quarks at $T=0$.}
\label{fig:free_MEM}
\end{figure}

\begin{figure}
\includegraphics[width=8.0cm]{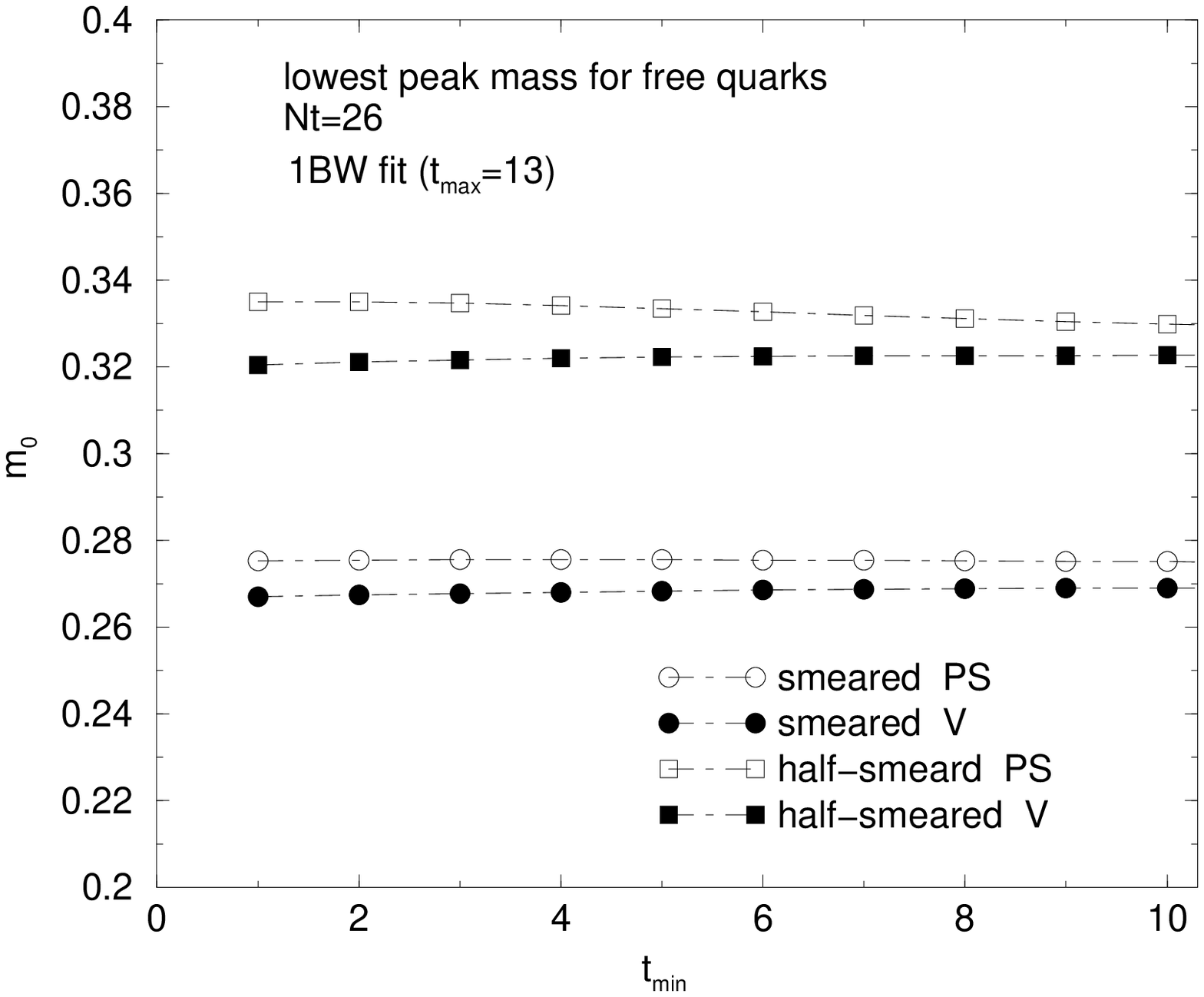}
\includegraphics[width=8.0cm]{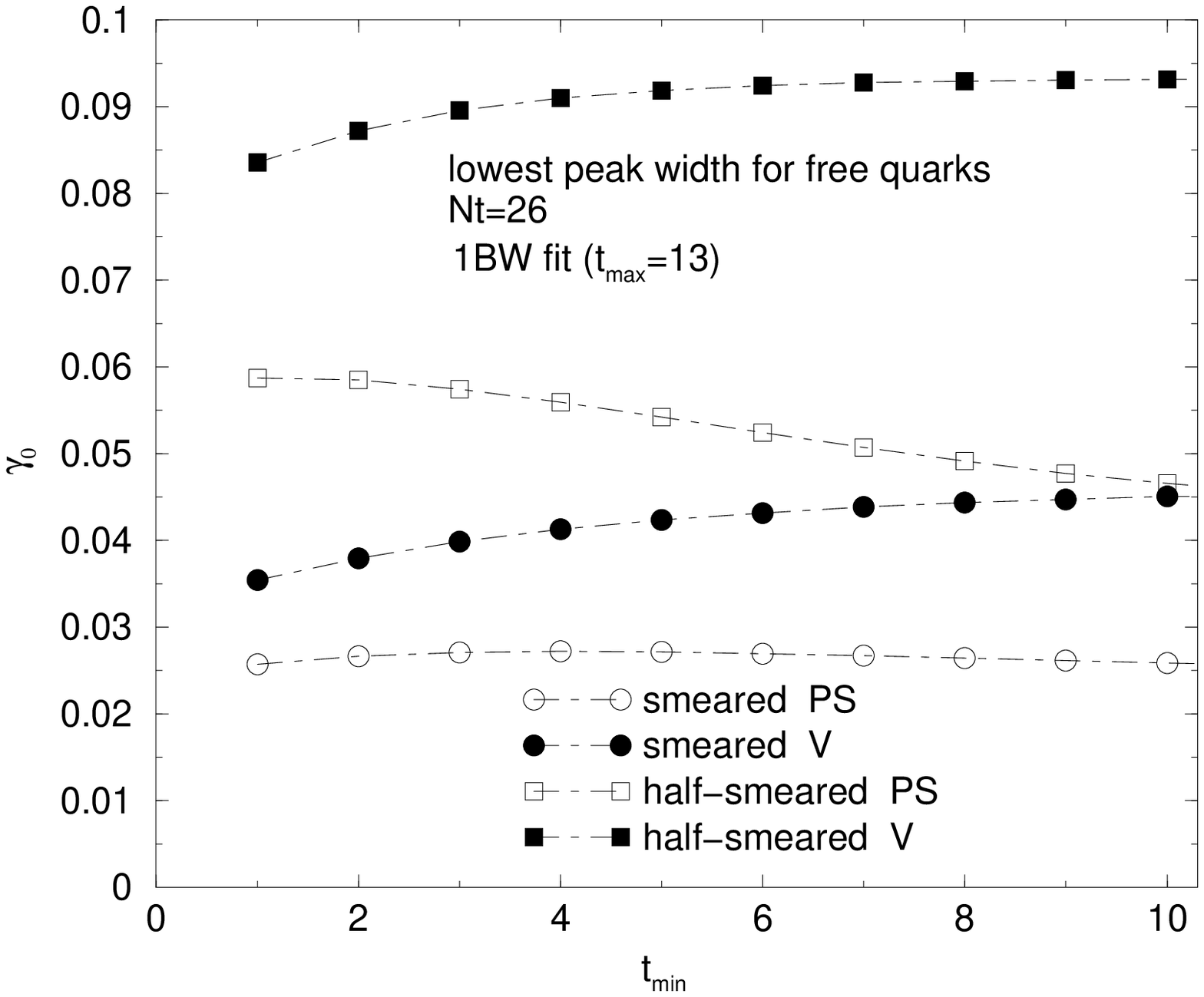}
\caption{The result of $\chi^2$ fit analysis for the spectral
functions from the PS correlators composed of free quarks
at $N_t=26$.}
\label{fig:free_fit}
\end{figure}

To learn how to distinguish physical features of the correlators from 
artifacts due to the smearing we apply
MEM and the $\chi^2$ fit analyses to the correlators composed of
free quarks on our lattice,
which were discussed in Sec.~\ref{subsec:free_quark}.

For the MEM analysis, the fluctuations of the correlators are given
by hand in the same manner as the mock data analyses
in Refs.~\cite{NAH99,CPPACS01a}.
The deviations of the correlators are less than those of the data
from the Monte Carlo simulation.
Figure~\ref{fig:free_MEM} shows the result of MEM applied to the
smeared correlator in pseudoscalar channel composed of free quarks
at $T=0$.
The upper bound of $t$ region used for the analysis, $t_{max}$,
is varied to observe how the result of MEM depends on the $t$-region
used.
With decreasing $t_{max}$, the resolution of the spectral function
becomes worse, and for $t_{max}\leq 16$ the spectral function
extracted with MEM displays just one peak.
Therefore for the circumstances specific at  $T>0$  MEM does not have 
sufficient resolution to distinguish individual states
if they are closer than about $ 0.05 a_\tau^{-1}$.
If the correlator is composed of almost free quarks, the width
extracted with MEM may be of order of 0.05--0.1 in temporal
lattice units.

We also apply the $\chi^2$ fit analysis to the correlators composed
of free quarks.
In this case the errors of the same size as in Monte Carlo simulation
are just put on the correlators without fluctuating them.
Figure~\ref{fig:free_fit} shows the result of the $\chi^2$ fit of
the correlators composed of free quarks at $N_t=26$ using
the 1-BW {\it ansatz}.
The $t_{min}$ dependence of mass and width (with fixed $t_{max}=13$)
indicates that the single BW {\it ansatz} describes rather
well the correlators.
As is observed in Figure~\ref{fig:free_fit},
if the correlator is composed of free quarks, 
 the $\chi^2$ fit gives sizable difference
in mass and width parameters for the smeared and half-smeared
correlators. This dependence is in agreement with
the fact that the propagator should only show the two free quark cut
and no particle-like excitations and indicates that by testing
the dependence of the result on the smearing function we can 
distinguish physical effects from artifacts due to smearing.

\section{Setup of numerical simulation}
 \label{sec:simulation}

\subsection{Lattice setup}

The zero temperature lattice used in this paper is the third one
of Ref.~\cite{Aniso01b}:
a quenched lattice of size $20^3\times 160$, generated with
the standard plaquette action with $(\beta, \gamma_G)=(6.10, 3.2108)$.
These coupling and bare anisotropy correspond to the
renormalized anisotropy $\xi=a_\sigma/a_\tau=4$ within
1\% accuracy \cite{Kla98}, and 
the spatial lattice cutoff $a_\sigma^{-1}=2.030(13)$ GeV set by the
hadronic radius $r_0$ \cite{Som94}.
At $T=0$, 500 configurations are generated with the pseudo-heat-bath
update algorithm, each separated by 2000 sweeps after 20000 sweeps for
thermalization.
The mean-field values are defined as the average values of link
variables in the Landau gauge, and obtained as 
$u_\sigma=0.8059(1)$ and $u_\tau=0.9901$.

To determine the critical temperature,
we measure the Polyakov loop susceptibility at $N_t=27$,
28, and 29 at $\beta=6.10$, and in addition,
at several values of $\beta$ (with corresponding values of $\gamma_G$) 
around $\beta=6.10$ at fixed $N_t=28$.
At $\beta=6.05$ the lattice scale set by $r_0$ is
$a_\sigma^{-1}=1.892(10)$ GeV, which together with $a_\sigma^{-1}$
at $\beta=6.10$ determine the scales at the other values of $\beta$
by linear interpolation.
The susceptibility peaks at about $\beta=6.10$ and $N_t=28$.
The critical temperature is obtained as $T_c\simeq 290$ MeV with
10 MeV of roughly estimated uncertainty.

The charmonium correlators at $T>0$ are measured for two values of
temporal lattice extent, $N_t=32$ and $26$.
Corresponding temperatures are 0.88$T_c$ for $N_t=32$,
and 1.08$T_c$ for $N_t=26$.
For brevity, these temperatures are hereafter referred to as
0.9$T_c$ and 1.1$T_c$, respectively.
Thus the temperatures treated in this paper are in the vicinity
of the transition.
At each of these two $N_t$'s, we generate 1000 configurations
each separated by 500 pseudo-heat-bath sweeps after 20000
sweeps for thermalization.

At $N_t=26$, we find that the configurations almost stay in
a single Polyakov loop sector during the whole updating history.
Transitions to other sectors occurred only once after
more than 470k sweeps.
In Ref.~\cite{TARO01} it was reported that the light mesonic
correlators behave differently on configurations in different
sectors.
We regard quenched lattices as approximations to lattices
with dynamical quarks, on which $\mathbf{Z}_3$ center symmetry is
explicitly broken and Polyakov loop prefers to stay on the real axis.
Therefore we transform all the configurations to the the real sector of the
Polyakov loop.

\subsection{Charmonium correlators}

As mentioned in Sec.~\ref{subsec:calibration},
we use $(\kappa, \gamma_F)=(0.112, 4.00)$, which correspond to
$m_Q\simeq 0.98$ GeV, roughly the charm quark mass.
The statistical errors for the results are estimated with
the jackknife method with appropriate binning.

In the next section, we start our analysis at $T=0$ with the examination
of the point correlators.
On each configuration, we calculate the point correlators four times
with four different source points:
$t=0$ and 80 at spatial site $\vec{x}=(0,0,0)$
and $t=40$ and 160 with $\vec{x}=(10,10,10)$.
Then these correlators are averaged with appropriate shifts in
$t$-direction.
At $T=0$ the standard $\chi^2$ fit of data to a multipole form works
well. To fix the general picture we anticipate on the analysis 
of the next sections and list
in Table~\ref{tab:spectrum}  the result of two-pole fits
of the point correlators in PS and V channels.
The fit ranges $[t_{min}, t_{max}]$ are determined by varying
$t_{min}$ with fixed $t_{max}=80$ and observing the stability
of the fit parameters.
The results of the ground state masses are very close to the
experimentally observed charmonium masses.
On the other hand, the hyperfine splitting is with about 74 MeV
smaller than the experimental value,
$m_{J/\psi}-m_{\eta_c} \simeq 117$ MeV.
This is a well-known feature of the quenched lattice simulations
and is considered mainly due to quenching,
although lattice artifacts in the charmonium system can also
play an important role \cite{TARO02}.

\begin{table}
\caption{
The spectrum at zero temperature determined from the point
and the smeared correlators.
The masses in PS and V channels are determined with the two-pole
fit.}
\begin{ruledtabular}
\begin{tabular}{cccc}
Correlator & State & $m_{PS}$  &  $m_V$ \\
\hline
Point   & ground      &  0.36835(37) & 0.37748(49) \\
        & first exc.  &  0.449(31) & 0.463(42) \\
        & fit range   & $45$--$80$ & $50$--$80$  \\
\hline
Smeared & ground      &  0.36856(9)& 0.37769(12) \\
        & first exc.  &  0.500(22) &  0.479(23)  \\
        & fit range   & $17$--$80$ & $15$--$80$  \\
\end{tabular}
\end{ruledtabular}
\label{tab:spectrum}
\end{table}

The smeared correlators are the main target of this paper
for reasons we already explained.
The smearing parameters are described
in Sec.~\ref{subsec:calibration}.
The result of a fit to 2-pole form is also shown in
Table~\ref{tab:spectrum}.
It is well-known that the correlators with smeared sink 
are quite noisy.
To reduce the noise, we measure sixteen correlators with different
source points on each configuration, and average them.
We change the spatial center of the smearing function, as well
as the time slice, to reduce the correlation as much as possible.
In contrast to a naive expectation, we find this procedure efficient.
In fact, at $T=0$, the statistical errors are reduced by a factor of
about 0.3 in the region $t=8$--16, which is the most important region
for the present analyses, in both the PS and V channels.
This reduction of errors corresponds to an increase in statistics of
about 10 times.
Almost the same amount of reduction of errors is observed at $N_t=32$.
At $N_t=26$, the errors are reduced by factors of 0.40--0.45 in the
range $t=8$--13.
This way of reducing statistical errors will be advantageous
in the case of dynamical simulations.

For the present kind of analysis, whether the analysis is efficient or
not may significantly depends on the statistics.
In the temporal region $t=8$--16, the statistical errors of
point correlators at $T=0$ are about 0.12--0.18\% and 0.08--0.15\%
for PS and V channels, respectively.
In the same region, the smeared correlators have statistical
errors of 0.14--0.17\% and 0.15--0.20\%.
Therefore at $T=0$ the errors are of the same order for the point
and smeared correlators.
At finite temperature, with 1000 configurations, the statistical
errors are of roughly the same size as at $T=0$.
We also note that the correlation between the correlators at
neighboring time slices is stronger for the point correlators than for
the smeared ones.

\begin{figure}
\vspace{-0.2cm}
\includegraphics[width=9.2cm]{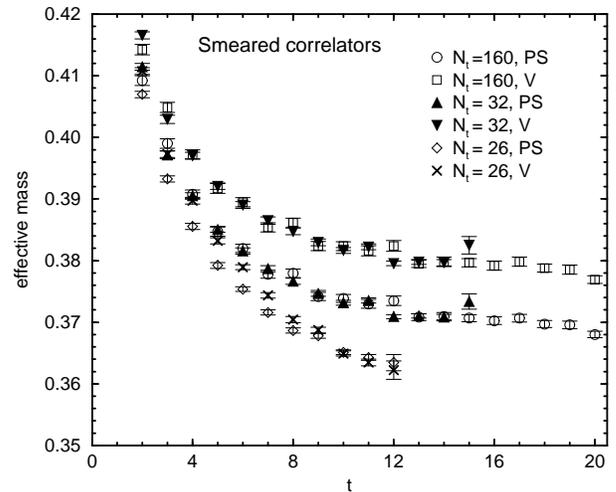}
\vspace{-0.9cm}
\caption{The effective masses for the smeared correlators
in PS and V channels at three temperatures.}
\label{fig:efp}
\end{figure}

To obtain an impression of
the physics to be expected at various temperatures we show
in Figure~\ref{fig:efp} the effective mass plot for the smeared
correlators, with $m_{eff}(t)$ defined through
\begin{equation}
 \frac{C(t)}{C(t+1)}
      = \frac{\cosh\left[m_{eff}(t)(N_t/2-t) \right]}
             {\cosh\left[m_{eff}(t)(N_t/2-t-1) \right]}.
\end{equation}
If the correlator is dominated by a single (stable) state,
$m_{eff}(t)$ shows a plateau. 
In Fig.~\ref{fig:efp}, the effective masses at $T=0$ show plateaus
beyond $t\simeq 16$.
The effective masses at $N_t=32$ ($T\simeq 0.9 T_c$) show almost the same behaviors
as at $T=0$.
This implies that the charmonium states are almost unchanged
at this temperature, compared to those at $T=0$.
In contrast, $m_{eff}(t)$'s at $N_t=26$ ($T\simeq 1.1 T_c$) show quite different
behavior:
they decrease rapidly compared to those at $T<T_c$, and in the large
$t$ region ($t \geq 10$) the pseudoscalar and vector channels are
almost degenerate.
Therefore a qualitatively different behavior of the spectral function
is expected below and above $T_c$.
These features are consistent with observations in earlier 
work \cite{Ume01}.

\section{Analysis at zero temperature}
\label{sec:result_zt}

\subsection{Result of MEM for point correlators}
 \label{subsec:MEMpoint}

We start with MEM analysis of the point correlators
at zero temperature.

First we discuss the default model function.
Following previous applications of MEM to lattice data
\cite{NAH99,CPPACS01a},
 we adopt as $m_{DM}$ in Eq.~(\ref{eq:default_model}) the asymptotic
values of correlators in the perturbative QCD,
\begin{equation}
 \frac{1}{\omega^2} A^{cont}(\omega \gg \Lambda_{QCD} )
  = \frac{r_1}{4\pi} \left( 1 + r_2 \frac{\alpha_s}{\pi} \right),
 \label{eq:corr_asymptotic}
\end{equation}
where $r_1=3/2$ and $r_2=11/3$ for PS channel, and 
$r_1=1$ and $r_2=1$ for V channel.
As value of the strong coupling constant $\alpha_s$,
we use $\alpha_s=0.2$ at 8 GeV from  Ref.~\cite{PDG00}
as a typical value at the temporal cutoff of the present lattice.
The matching of the lattice theory with the continuum theory is
performed through the tadpole-improved tree-level, and hence
the renormalization of the quark field is represented only by
the KLM normalization factor (\ref{eq:KLM_factor}).
These settings give for the parameter of the default model function
$m_{DM}=4.1$ for PS channel, and 2.4 for V channel.
If MEM works as a method to extract the spectral
function without assuming a specific form,
the result should not be sensitive to the choice of this
parameter.
In addition, the observed asymptotic behavior of the spectral function
from lattice data around the cutoff is actually different from that of
Eq.~(\ref{eq:corr_asymptotic}).
We therefore observe the dependence of the result on the value of
$m_{DM}$ by changing $m_{DM}$ by factors of $10$ and $0.1$, to verify
the insensitivity of the result to $m_{DM}$.

The further setting of MEM is as follows.
In all cases, the correlator at $t=0$ is not used for the analysis.
The minimum of $t$ used depends on the type of analysis, 
in most cases $t_{min}=1$ is adopted.
The frequency $\omega$ is discretized with $\Delta\omega=0.002$,
in the region $[ \omega_{min}, \omega_{max} ]=[0.001,4.0]$.
The dependence of the result on the parameters for $\omega$
is sufficiently small around the adopted values.

\begin{figure}
\vspace{0.1cm}
\includegraphics[width=8.0cm]{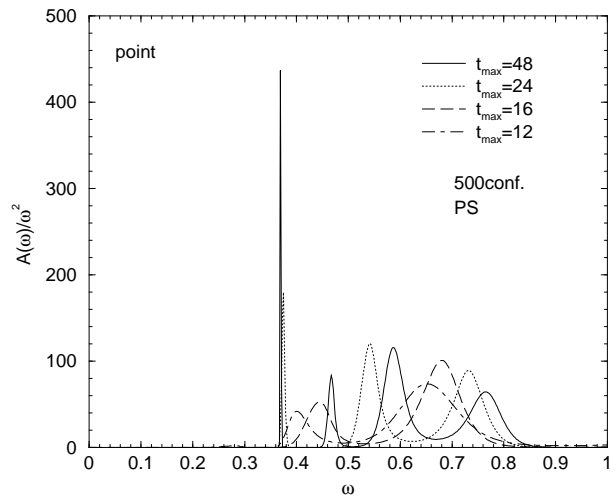}
\caption{
The result of MEM analysis of the point correlator in PS
channel at $T=0$.
The figure shows the $t_{max}$ dependence of the resultant
spectral function.}
\label{fig:MEM_zt_point1}
\end{figure}

\begin{figure}
\vspace{0.1cm}
\includegraphics[width=8.0cm]{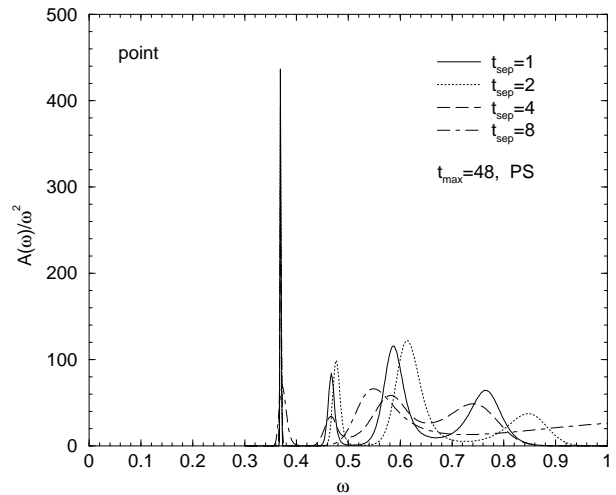}
\caption{
The result of MEM analysis of the point correlator in PS
channel at $T=0$.
This figure shows the $t_{sep}$ dependence of the result
of the spectral function with fixed $t_{max}=48$.}
\label{fig:MEM_zt_point2}
\end{figure}

We first examine how the restriction of degrees of freedom of the
correlator affects the reconstructed spectral function.
Three parameters $t_{min}$, $t_{max}$ and $t_{sep}$ are introduced,
that is MEM is applied for the correlators at times
$t_{min} \leq t \leq t_{max}$ in every $t_{sep}$ time slices.
We examine two types of restrictions of the numbers of degrees of
freedom:
\begin{description}
 \item[\rm (I).] $t_{max}$ is varied as 48, 24, 16, and 12, 
   with fixed $t_{min}=1$ and $t_{sep}=1$.
   The corresponding numbers of data points are the same as $t_{max}$.
 \item[\rm (II).] $t_{sep}$ is varied as 1, 2, 4, and 8, with fixed
   $t_{max}=48$.
   $t_{min}$ is varied accordingly as $t_{min}=t_{sep}$.
   Correspondingly, the numbers of data points are 48, 24, 12, and 6,
   respectively.
\end{description}
From the point of view of the basis functions of  the singular value decomposition
in the spectral function space,
case (I) reduces the number of basis functions while keeping the
functions unchanged.
On the other hand, case (II) dilate the basis functions
by a factor of $t_{sep}$ simultaneously reducing their number.
Reliability of the result at finite temperature requires the obtained spectral
function to be stable under the type (I) restriction.
Although larger values for $t_{max}$ are possible, $t_{max}=48$
is sufficiently large and has appropriate physical range
($t_{max}a_\tau\simeq 1.2$ fm) for the present purpose.

The result is shown in Figs.~\ref{fig:MEM_zt_point1}
and \ref{fig:MEM_zt_point2} for the pseudoscalar channel.
For the vector correlator a similar result is obtained.
In the case of $t_{max}=48$ and $t_{sep}=1$, i.e.
without restriction, the reconstruction of the spectral functions
seems successful, and their fundamental features
appear to be the same as in previous works \cite{NAH99,CPPACS01a}.
The peak positions of the ground and first excited states are
consistent with the result of $\chi^2$ fit summarized
in Table~\ref{tab:spectrum}.
Here we do not discuss the higher excited states because of
uncertainties not only in MEM but also in the $\chi^2$ fit analysis.

Figure~\ref{fig:MEM_zt_point1} displays the
result under restriction conditions (I).
Decreasing $t_{max}$, the spectral function becomes increasingly
different from that with $t_{max}=48$.
For $t_{max}\leq 24$, the first excited state peak does not appear
at the correct place,
and for $t_{max}\leq 16$ even the ground state peak is located at
an incorrect place and has broad width.
These lowest peaks for different $t_{max}$'s are all significant
in the sense of the error analysis of MEM \cite{NAH99}.
Therefore $t_{max}\leq 16$ is not acceptable for extracting reliable
information of the ground state from the point correlators.
Since at finite temperature we are restricted to $t_{max}\leq 16$
because of short temporal extent this
is the reason why we abandon using the point correlators
and apply the smearing technique.

Figure~\ref{fig:MEM_zt_point2} shows the
result with restriction condition (II).
In this case, the reconstructed spectral functions are rather stable,
while the sharpness of peaks is lost by increasing $t_{sep}$.
Even with 6 degrees of freedom (as a linear combination of 6
functions), the reconstructed spectral function at least exhibits
the ground state peak at the correct position.
To summarize, if one has a temporal region of the correlator of the order
of 1 fm, the spectral function can be rather nicely reconstructed
even with less than 10 degrees of freedom.

From these observation, we conclude that
with $O(10)$ degrees of freedom the
crucial condition is determined by the region of
$t$ where the correlators are measured.
$t_{max}a_\tau$ of the order of 1 fm seems necessary for a reliable
extraction of the spectral function from the point correlator.
Increasing the number of degrees of freedom above $10$ with fixed physical
range of $[t_{mim}, t_{max}]$ improves the situation only slightly.

Now  we consider the effect of  the statistics.
We apply the same analysis of the case (I) restriction to
the correlators averaged over first 100 configurations.
The observed features of  $t_{max}$ dependence are essentially the same
as with 500 configurations.
Therefore at the present level of statistics (order of several hundreds),
increasing statistics does not seem to improve drastically
the situation of the $t_{max}$ dependence.
We also compare the correlators averaged over 100
configurations randomly selected from the total of 500 configurations.
A comparison of the results with $t_{max}=48$ from five such correlators
shows that the first excited state does not always appear at the
same place as with 500 configurations, but deviates in a
range of $\omega=0.44$--0.52.
Thus with less statistics only the ground state
peak can be trusted even with $t_{max}=48$.
These observations indicate the 
required statistics for each situation.

Finally we briefly discuss the dependence on the default model.
We compare the results obtained with $m_{DM}$ in
Eq.~(\ref{eq:default_model})  multiplied by 10 and 0.1
with otherwise fixed parameters, $t_{max}=48$ and $t_{sep}=1$.
These results of the spectral function display the peaks at the same
places as with original $m_{DM}$, while increasing $m_{DM}$
the peaks become sharper.
Although present analysis is quite simple, we conclude that
the default model dependence of the result of MEM is not severe,
at least for a rough estimate of the shape of the spectral function.

The most important conclusion in the analysis of the point correlator
is that we need the order of 1 fm for the range of $t$ where the
correlators are measured, for reliable extraction of the spectral
function from the correlator with present level of statistics.
This requirement cannot be fulfilled at finite temperature,
and a brute force application of MEM to the point correlators
at $T>0$ would produce an unreliable result.
This result warns us against physical significance of the results
in Refs.~\cite{Bielefeld02a,Bielefeld02b,Bielefeld02c,AHN02}.
This difficulty may originate in that the point correlator
at short distance contains the contribution from wide range of
frequency up to the lattice cutoff.
We therefore give up to analyze further the point correlators,
and concentrate our attention on the smeared correlators.

\subsection{Result of MEM for smeared correlators}
 \label{subsec:MEMsmeared}

\begin{figure}
\vspace{0.1cm}
\includegraphics[width=8.0cm]{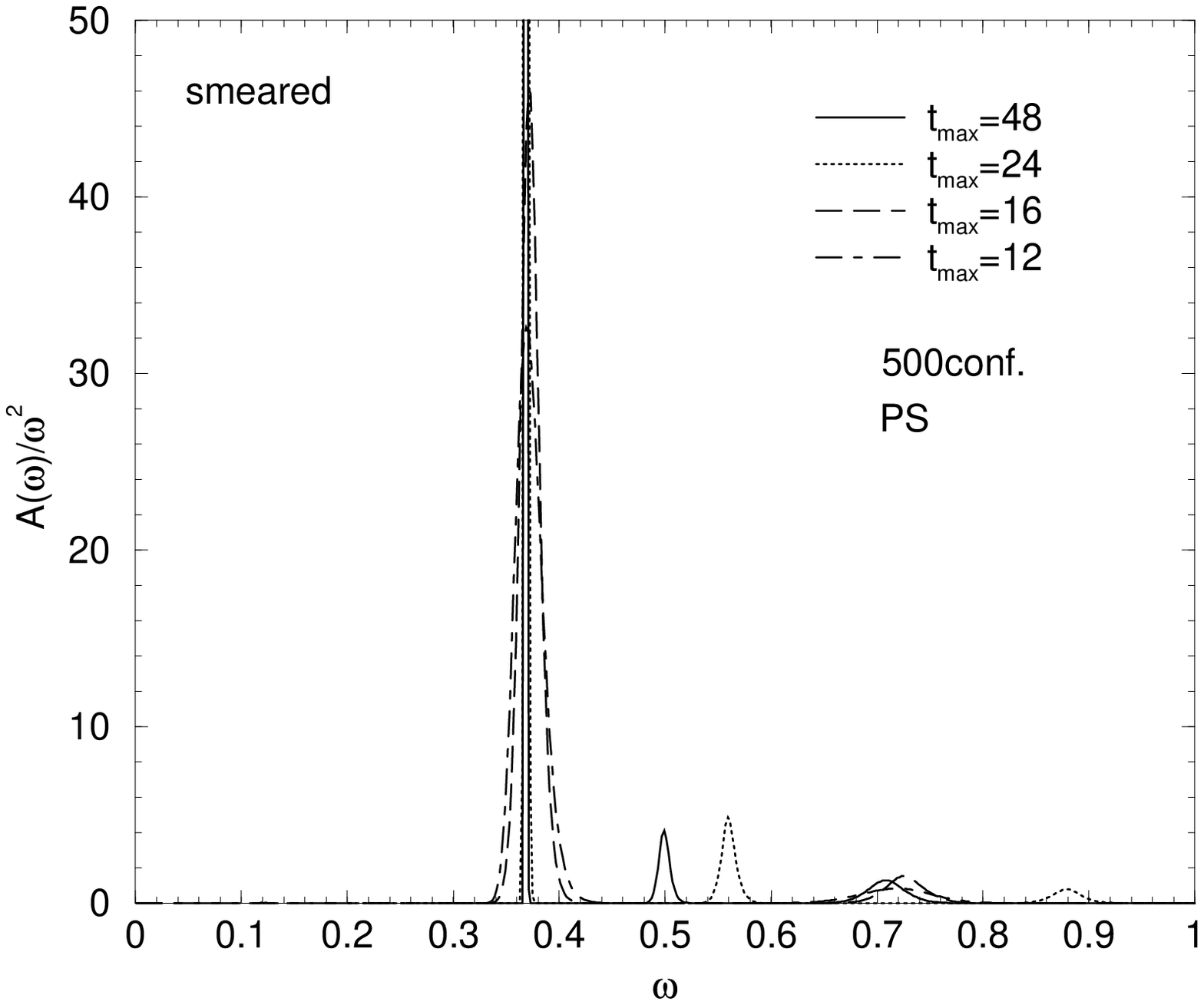}
\includegraphics[width=8.0cm]{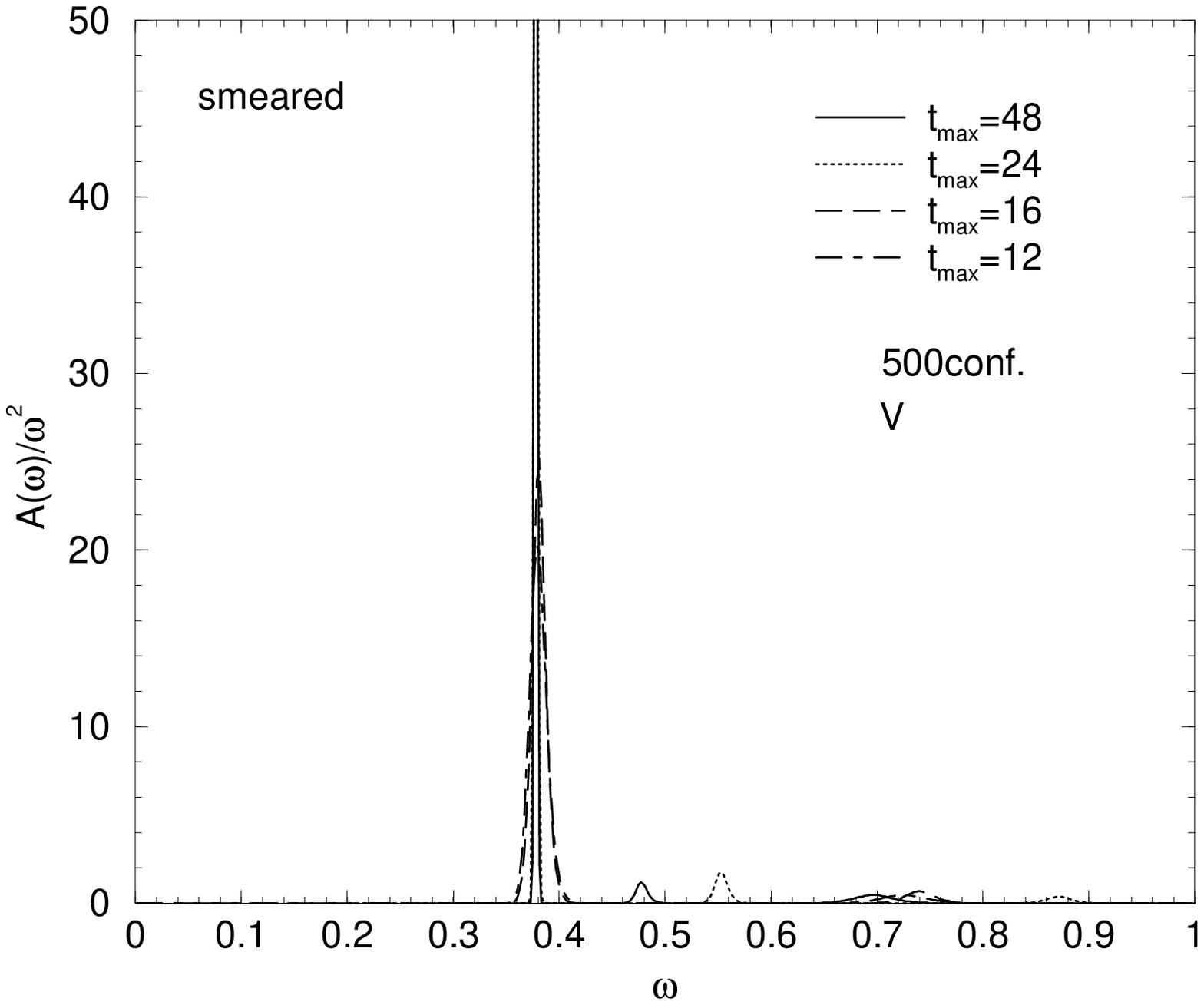}
\caption{
The spectral functions at $N_t=160$ determined with MEM.
The top and bottom panels show the $t_{max}$ dependences
of obtained spectral functions in PS and V channels, respectively.}
\label{fig:mem_160}
\end{figure}

\begin{figure}
\vspace{0.1cm}
\includegraphics[width=8.0cm]{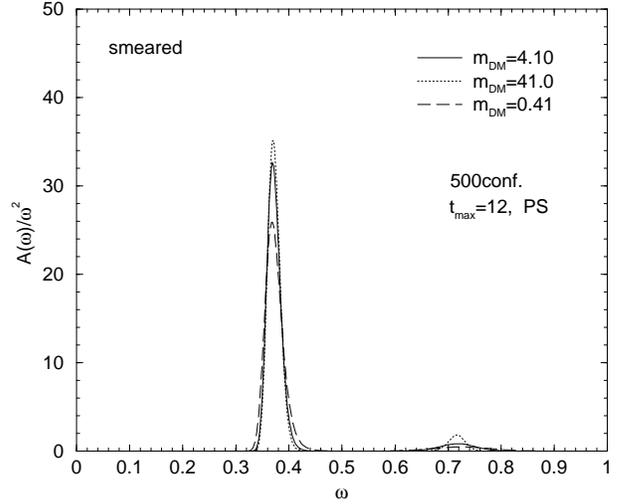}
\caption{
The default model dependence of the extracted spectral function
with MEM from the smeared correlators in PS channel.
The figure shows the result of MEM with three values of
$m_{DM}$, the parameter of the default model function
Eq.~(\ref{eq:default_model}).}
\label{fig:mem_160b}
\end{figure}

We now turn to MEM analysis of the smeared correlators.
Since the smearing technique reduces the high frequency part,
there is no a priori choice for the form of the default model.
As described in Sec.~\ref{subsec:procedure_MEM},
we use the same default model function as for the
point correlator, with the normalization of the smeared correlator
which gives the same overlap with the ground state as
the point correlator.
The default model dependence is examined in the same manner
as for the point correlators.
Other parameters are the same as for the point correlator,
except that $\omega_{max}=2.0$ is adopted, since the high frequency
region is sufficiently suppressed.

First we examine the dependence of the result on the numbers of
degrees of freedom, using the same restriction conditions
as for the point correlators.
Since the smearing technique enhances the low energy part,
we only consider whether the ground state peak is
correctly reproduced or not.
The result for the restriction condition (I) is displayed
in Fig.~\ref{fig:mem_160}.
In the both cases of PS and V channels  the place of the ground state
peak is stable with decreasing $t_{max}$, although the width is
gradually broadened.
Since the ground state peak evidently has no physical width at $T=0$,
this fictitious width is a systematic uncertainty of MEM due to the
insufficient number of  basis functions in the spectral function
space.
Therefore, in particular at finite temperature, an estimate of the
physical width requires a careful analysis combined with other
procedures.
The instability of  the first excited state peak under
changing $t_{max}$
is explained by the smallness of the contribution to the correlators
from energies above the ground state.
In fact, the two-pole fit analysis indicates that the overlap
of the first excited state with the smeared correlator is
about 7\% (6\%) of that of the ground state for the PS (V) channel.
Thus the instability of the excited state peaks has no
 significance for the following analysis.
We conclude that for the smeared correlators, MEM with
$t_{max}\simeq 12$ would work for a rough estimate of the structure
of the spectral function in the low energy part, such as examination of
whether strong ground state peak exist, while a quantitative
evaluation of width, for example, is difficult.
In the case of the restriction condition (II), a similar
tendency as for the point correlators is observed:
the position of the ground state peak is stable with increasing $t_{sep}$,
while the sharpness of the peak is gradually lost.

The $m_{DM}$ dependence of the result is examined in two cases of
$t_{max}$, $t_{max}=48$ and $12$, by using rescaling factors
10 and 0.1.
In the case of $t_{max}=48$, the ground state peak is strong,
and changing $m_{DM}$ does not cause a large effect other than the
decrease of the peak height with slight increase of the width
when decreasing $m_{DM}$.
Fig.~\ref{fig:mem_160b} shows the case of $t_{max}=12$
for the PS channel.
Although the peak shape is broadened with decreasing $m_{DM}$,
the essential features of the result are stable.
A similar result is observed for the V channel.

Let us summarize the  MEM analysis of the smeared correlators at $T=0$.
In contrast to the case of the point correlators,
MEM with restricted numbers of degrees of freedom works up to
$t_{max}=12$, i.e.  $t_{max}a_\tau \simeq 0.3$ fm, 
at least for a rough estimate of the shape of the spectral function
in the low energy part.
For this purpose, systematic uncertainties in the default
model parameter do not appear  significant.
This is an encouraging result for an application of MEM to finite
temperature as a precedent analysis to the $\chi^2$ fit.

\subsection{Result of $\chi^2$ fit analysis}

\begin{figure}
\vspace{-0.4cm}
\includegraphics[width=9.0cm]{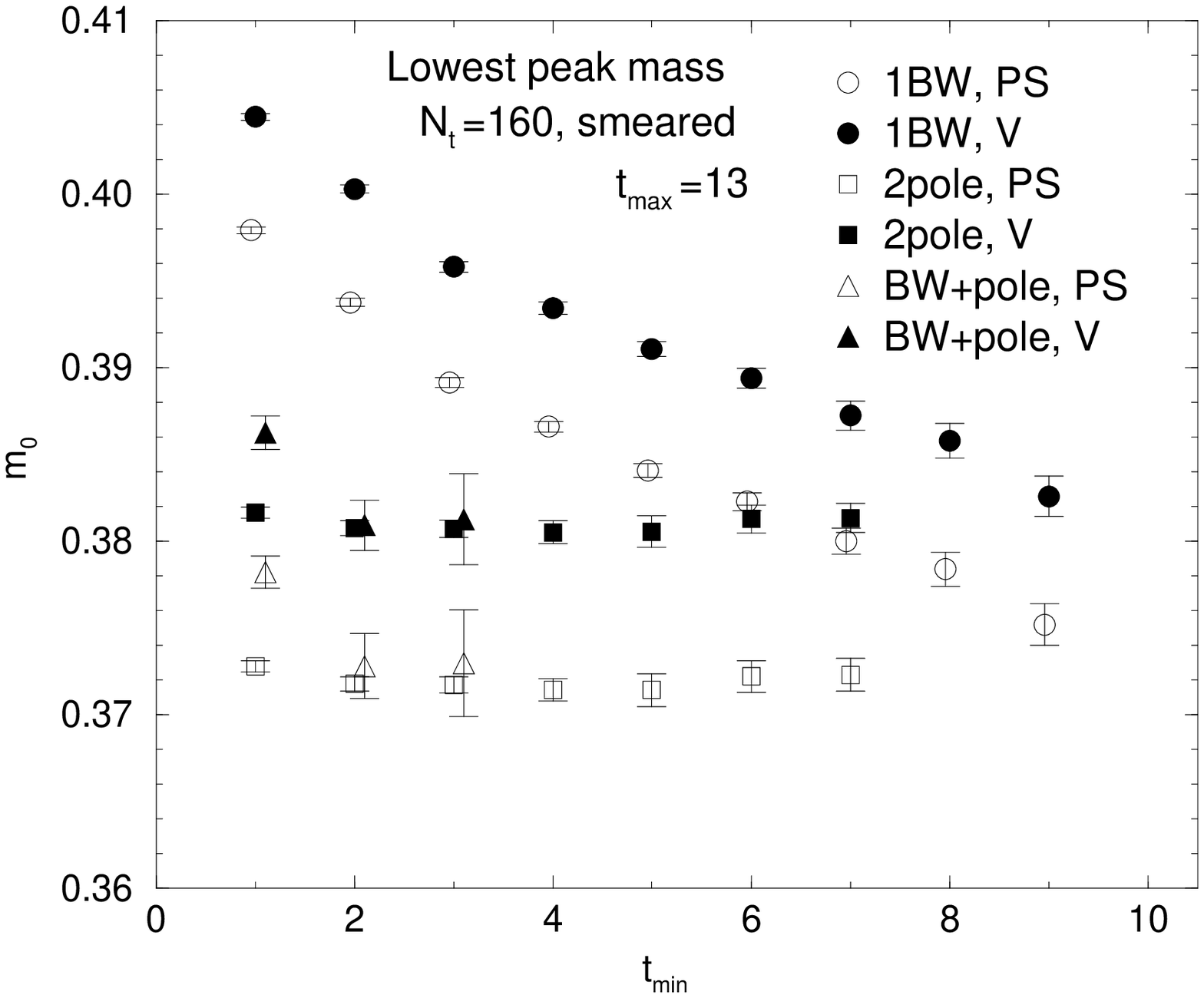}
\vspace{-1.2cm}\\
\includegraphics[width=9.0cm]{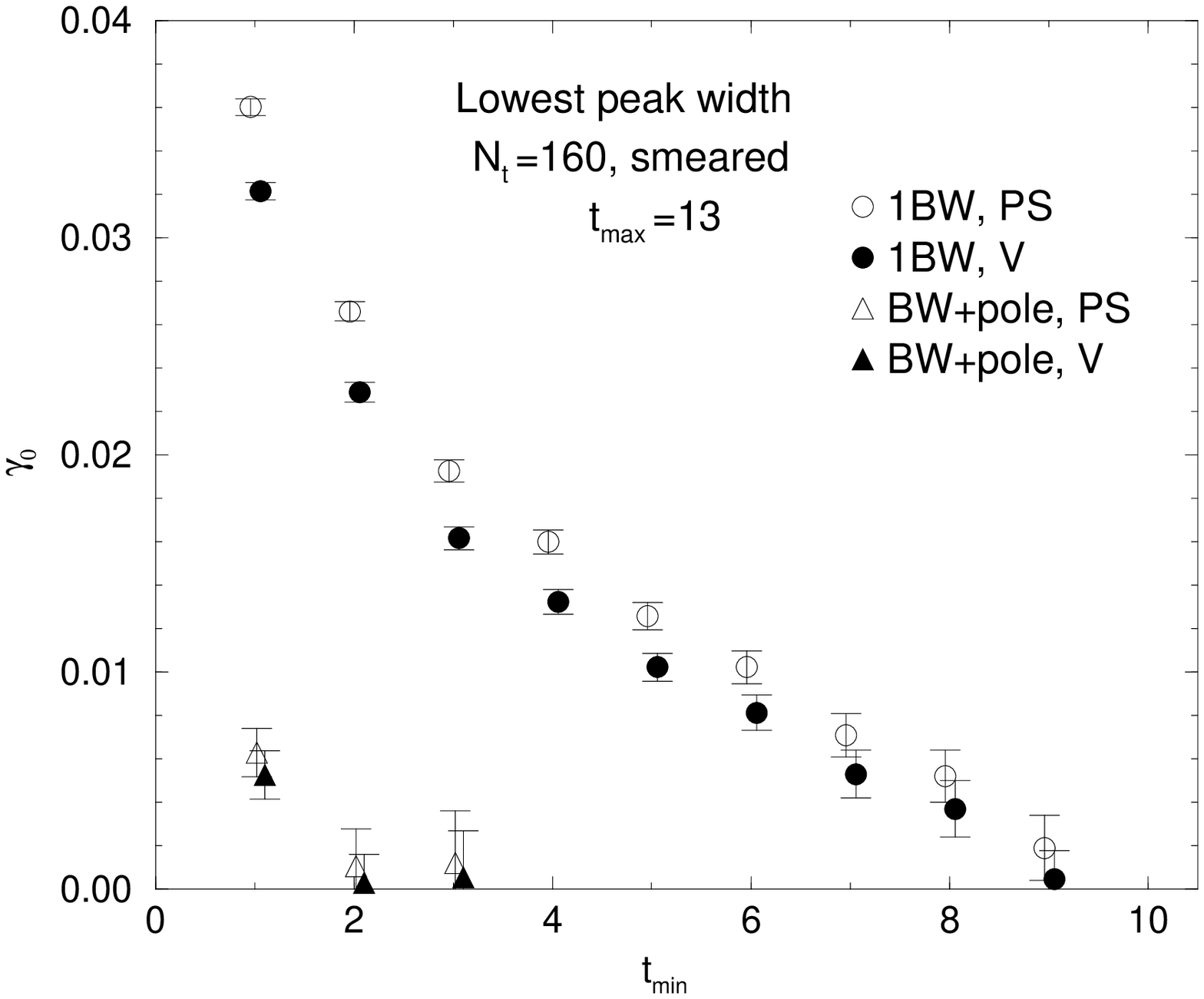}
\vspace{-0.8cm}\\
\caption{
The result of the $\chi^2$ fit analysis at $N_t=160$.
The top and bottom panels show the results for  mass and width
of the ground state peak, respectively.}
\label{fig:fit_160}
\end{figure}

The results with 2-pole {\it ansatz} have been given
in Table~\ref{tab:spectrum}.
For completeness, we also apply the $\chi^2$ fits to the smeared
correlators with forms other than the 2-pole fit.
The result is shown in Fig.~\ref{fig:fit_160}.
As fit range $[t_{min}, t_{max}]$
we fix $t_{max}=13$, considering the severest case at $T>0$,
and observe the $t_{min}$ dependence of the result.
The top and bottom panels of Fig.~\ref{fig:fit_160} show the
results of mass and width of the ground state peak, respectively.

As is expected, the 2-pole form describes well the correlators, and the
other two fit forms are consistent with the 2-pole fit.
In the cases of 1-BW and BW+pole fits  the $t_{min}$ dependences
of the widths are significant.
In the bottom panel the results for the width decreases as $t_{min}$
increases, and seem to approach zero.
Therefore no indication of a finite width is found at $T=0$,
as it should be.

\section{Analysis at $T<T_c$}
\label{sec:result_ft1}

Now the two analysis procedures are applied to
the correlators at $T\simeq 0.9 T_c$ ($N_t=32$).
In this and the next sections we no longer discuss the point
correlators and focus only on the smeared ones.

\subsection{Result of MEM analysis}

\begin{figure}
\vspace{0.1cm}
\includegraphics[width=8.0cm]{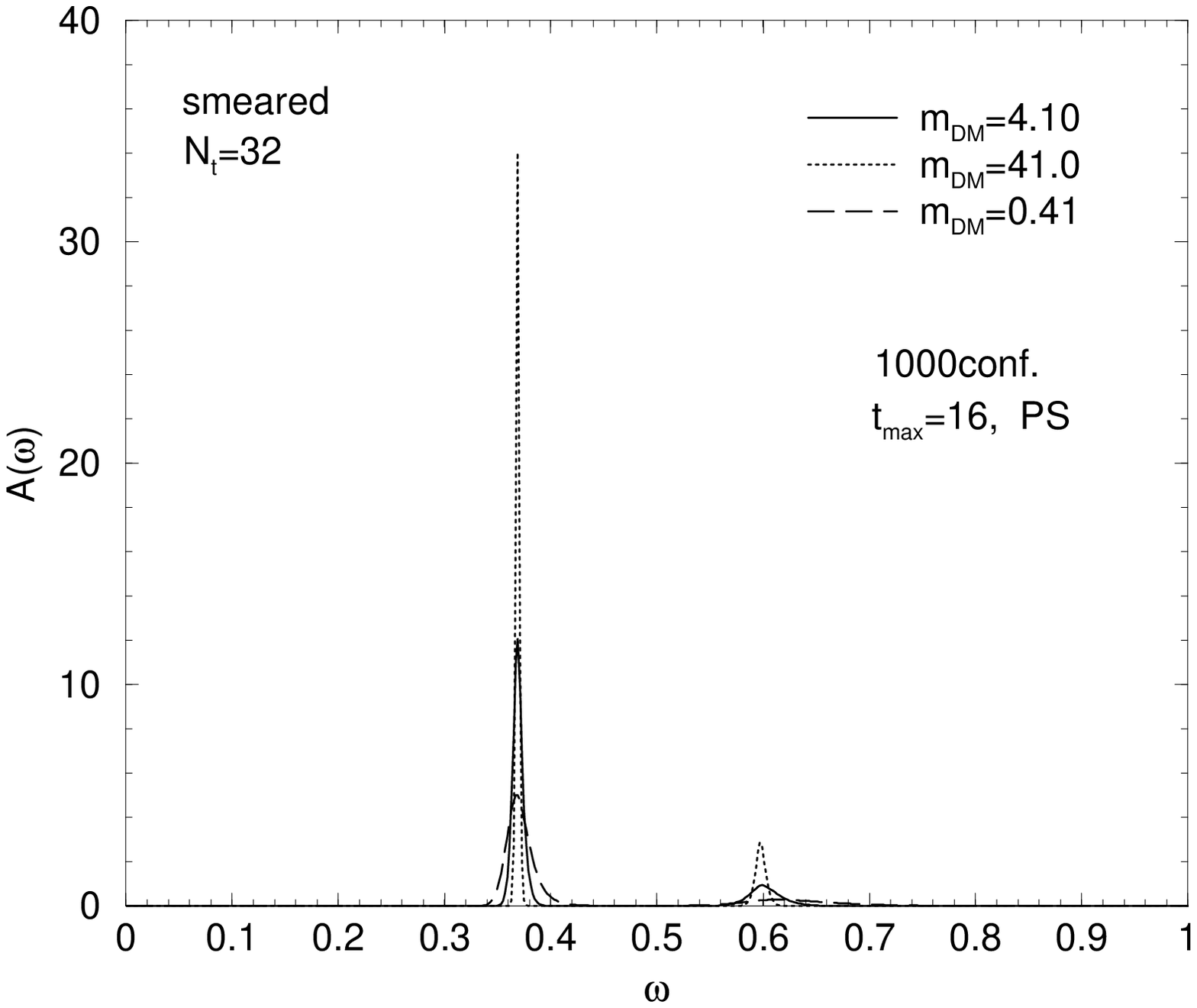}
\includegraphics[width=8.0cm]{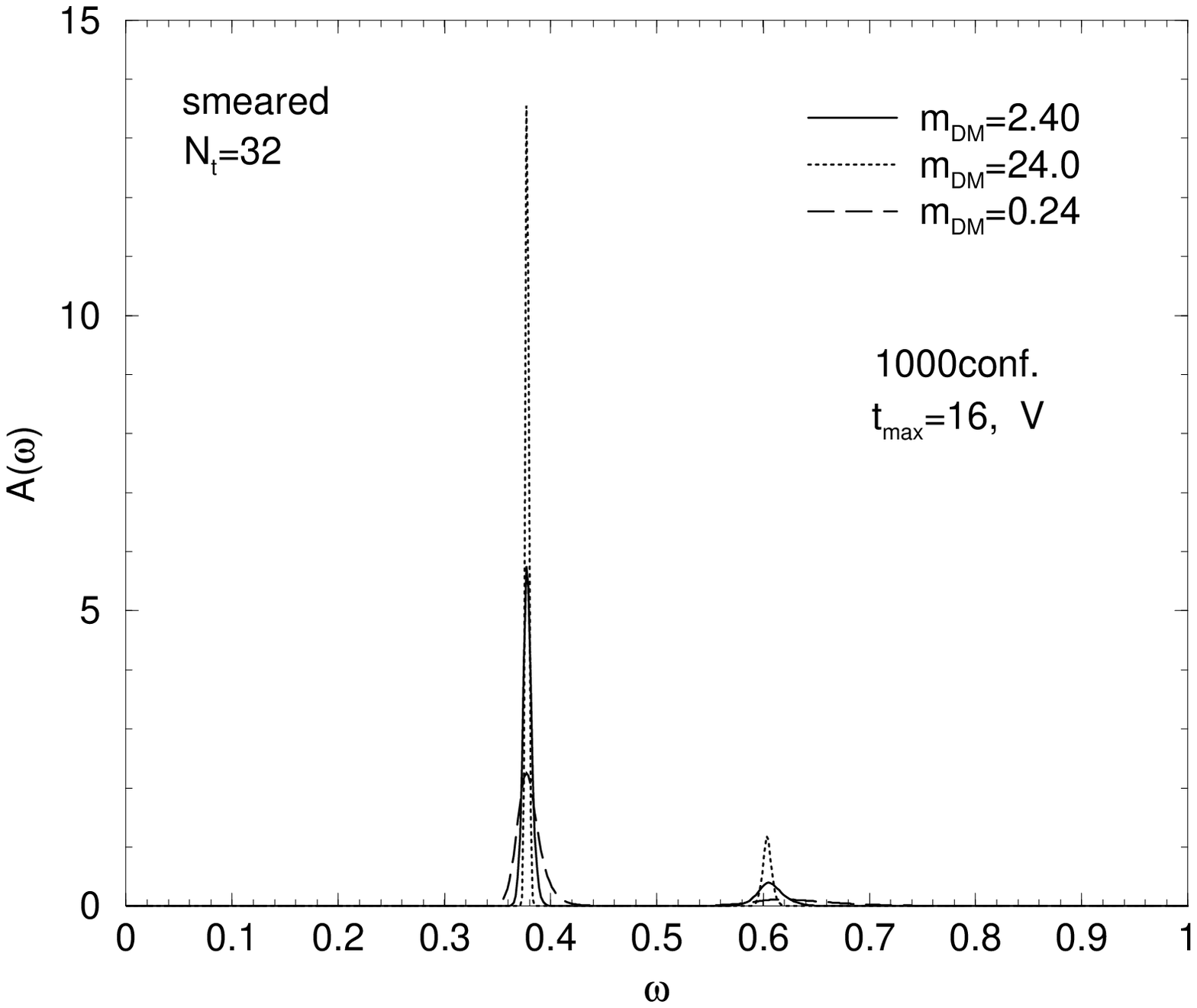}
\caption{
Spectral functions at $N_t=32$ determined with MEM.}
\label{fig:mem_32}
\end{figure}

We begin with the MEM analysis, in accordance with our strategy.
The parameters concerning MEM are almost the same as for
the smeared correlators at $T=0$.
The range of $t$ is fixed to $t_{min}=1$,
$t_{sep}=1$, and $t_{max}=16$.

The result is displayed in Fig.~\ref{fig:mem_32}
for three values of $m_{DM}$.
For the reasons already mentioned we focus on the ground state
peak and do not discuss in detail the high frequency part of
the spectral function.
As apparent in the figure a ground state peak appears in both
the PS and V channels.
Although with decreasing  $m_{DM}$ the widths of the ground state
peaks increase, the positions of the peaks are stable and almost the same
as at $T=0$.
In the present case, since there is no intrinsically advisable value
for $m_{DM}$ beyond an order estimate,
this ambiguity of the width of the peak should be considered
as an uncertainty of MEM applied to the smeared correlators.
We also perform the same analysis with less statistics, 500
configurations, to see how this result depends on the statistics.
The result is essentially the same, and thus statistically stable.

These results support the assumption that the mesonic ground states are
persistent up to this temperature, with almost the same masses
as at $T=0$.
The width of the ground state peaks is small while finite,
although it strongly depends on the value of $m_{DM}$.
Whether the width is physically finite or not should therefore be examined
with the $\chi^2$ fit analysis.

\subsection{Result of $\chi^2$ fit analysis}

\begin{figure}
\vspace{-0.4cm}
\includegraphics[width=9.0cm]{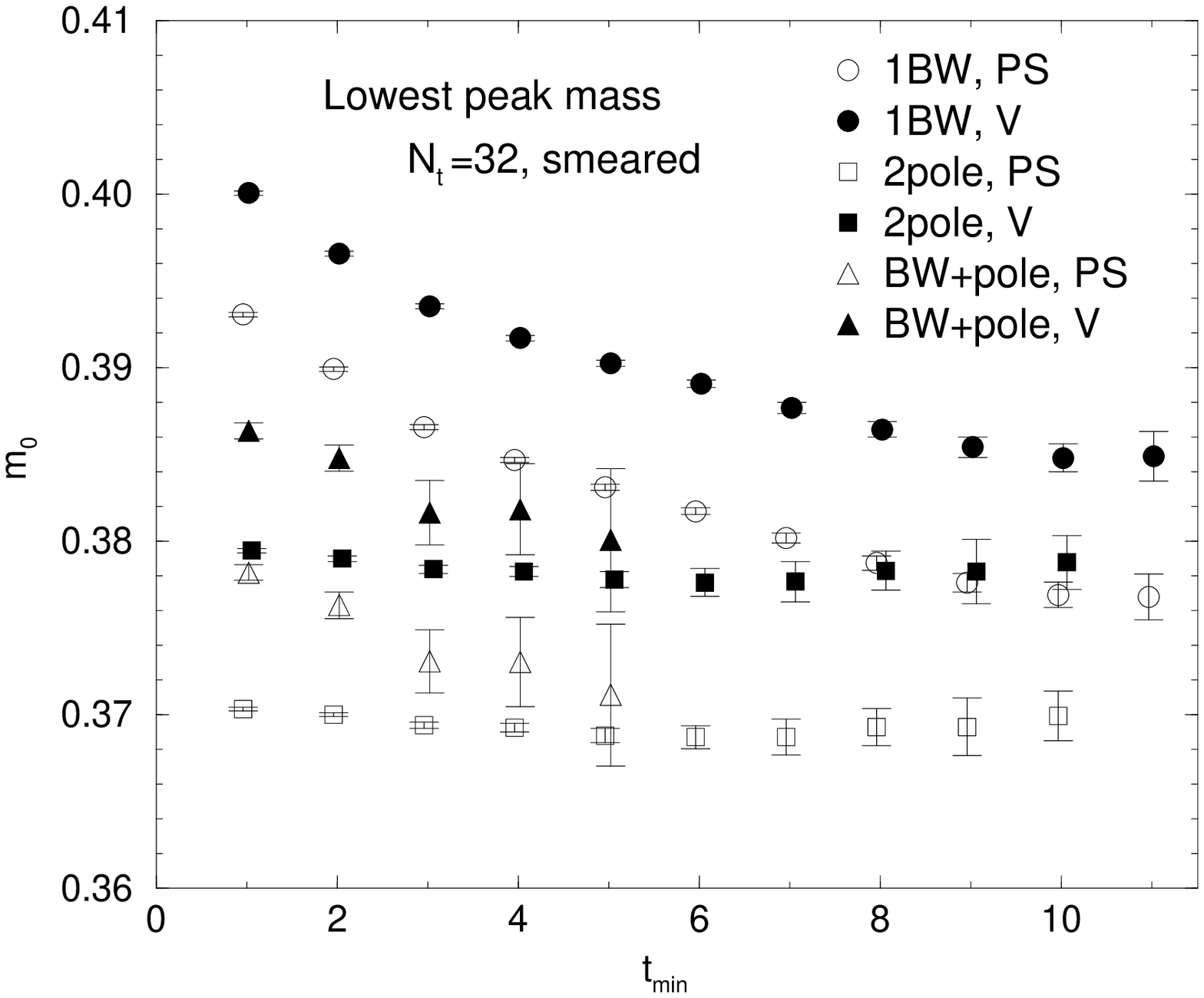}
\vspace{-1.2cm}\\
\includegraphics[width=9.0cm]{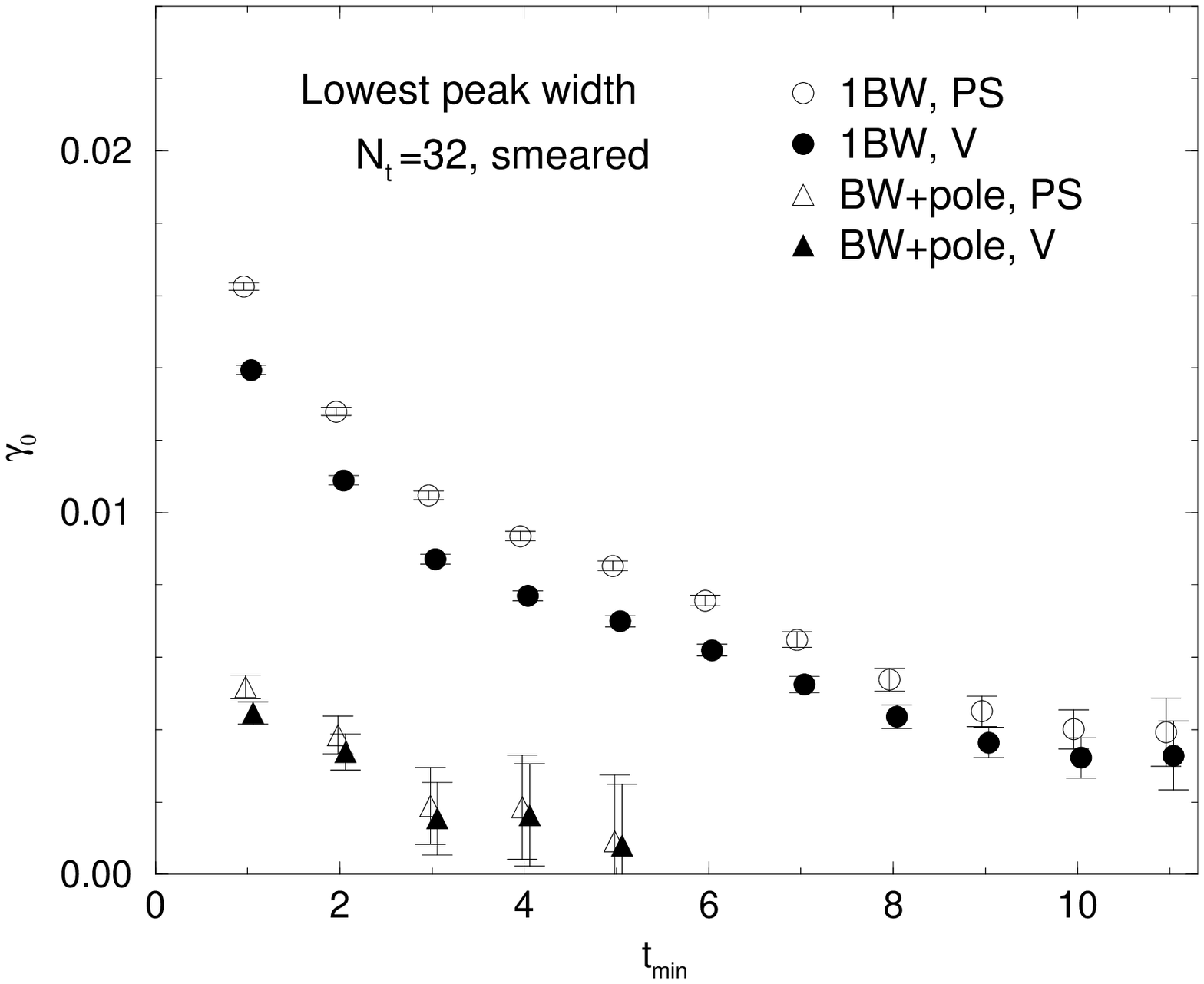}
\vspace{-0.8cm}
\caption{
The result of $\chi^2$ fit analysis at $N_t=32$.
The top and bottom panels show the results of the mass and width
parameters for the ground state peak, respectively.}
\label{fig:fit_32}
\end{figure}

According to the result of MEM we can assume that the spectral
function of the smeared correlator at this temperature
has a peak structure at almost the same mass as at $T=0$ and
with small width.
Therefore it is expected that the $\chi^2$ fits with the three types
of fit forms, 2-pole, 1-BW, and BW+pole forms can clarify
the low energy structure of the spectral function.

The results are summarized in Fig.~\ref{fig:fit_32}.
The top panel displays the dependence of the mass parameters
for the ground state peak on the lower bound of the fit range,
$t_{min}$.
The upper bound is fixed to $t_{max}=16$.
It is apparent that the fit to the 2-pole form exhibits stable
results both for the PS and V channels beyond $t_{min}=3$.
The result of fit to 1-BW form exhibits no plateau, indicating
that this form does not explain the whole structure of the correlators.
However, the values seem to approach to the corresponding results
of the 2-pole fits.
The result of fit to the BW+pole form is consistent with those
of 2-pole fit at $t_{min}\leq 2$.

The bottom panel of Fig.~\ref{fig:fit_32} shows the result for
the width parameter.
In the case of the fit to the 1-BW form the value of width
gradually decreases with  increasing $t_{min}$.
This behavior is consistent with a vanishing width.
The fit to BW+pole form also indicate that the width
is consistent with zero.
Therefore, there is no indication of a finite width for the spectral
function at this temperature.

All the results of fits to the three forms indicate
that the ground state peak is well described by a strong peak
with vanishing width, i.e. a pole-like structure,
and the associated mass is almost the same as at zero temperature.

\section{Analysis at $T>T_c$}
\label{sec:result_ft2}

\subsection{Result of MEM analysis}

\begin{figure}
\vspace{0.1cm}
\includegraphics[width=8.1cm]{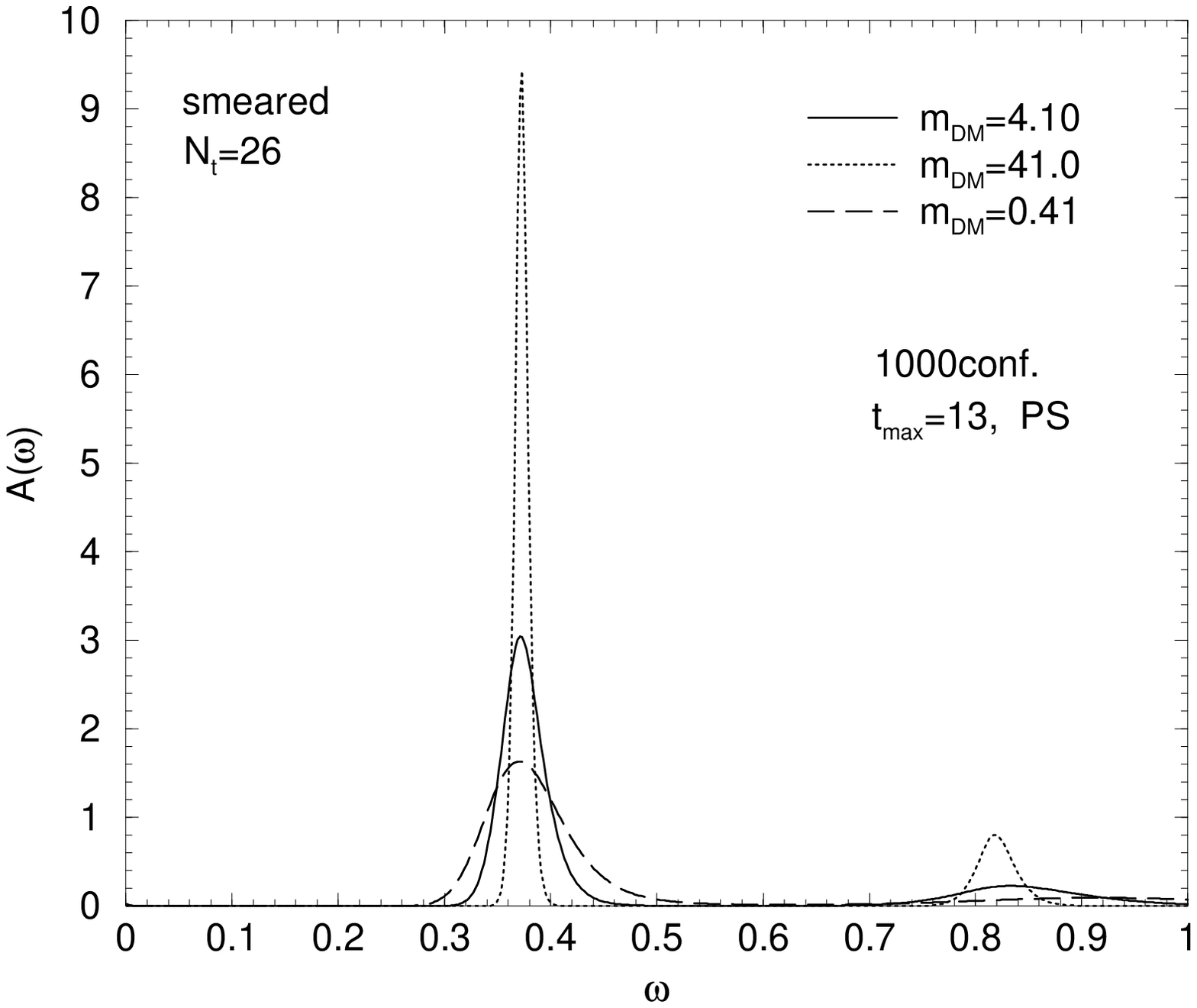}
\includegraphics[width=8.0cm]{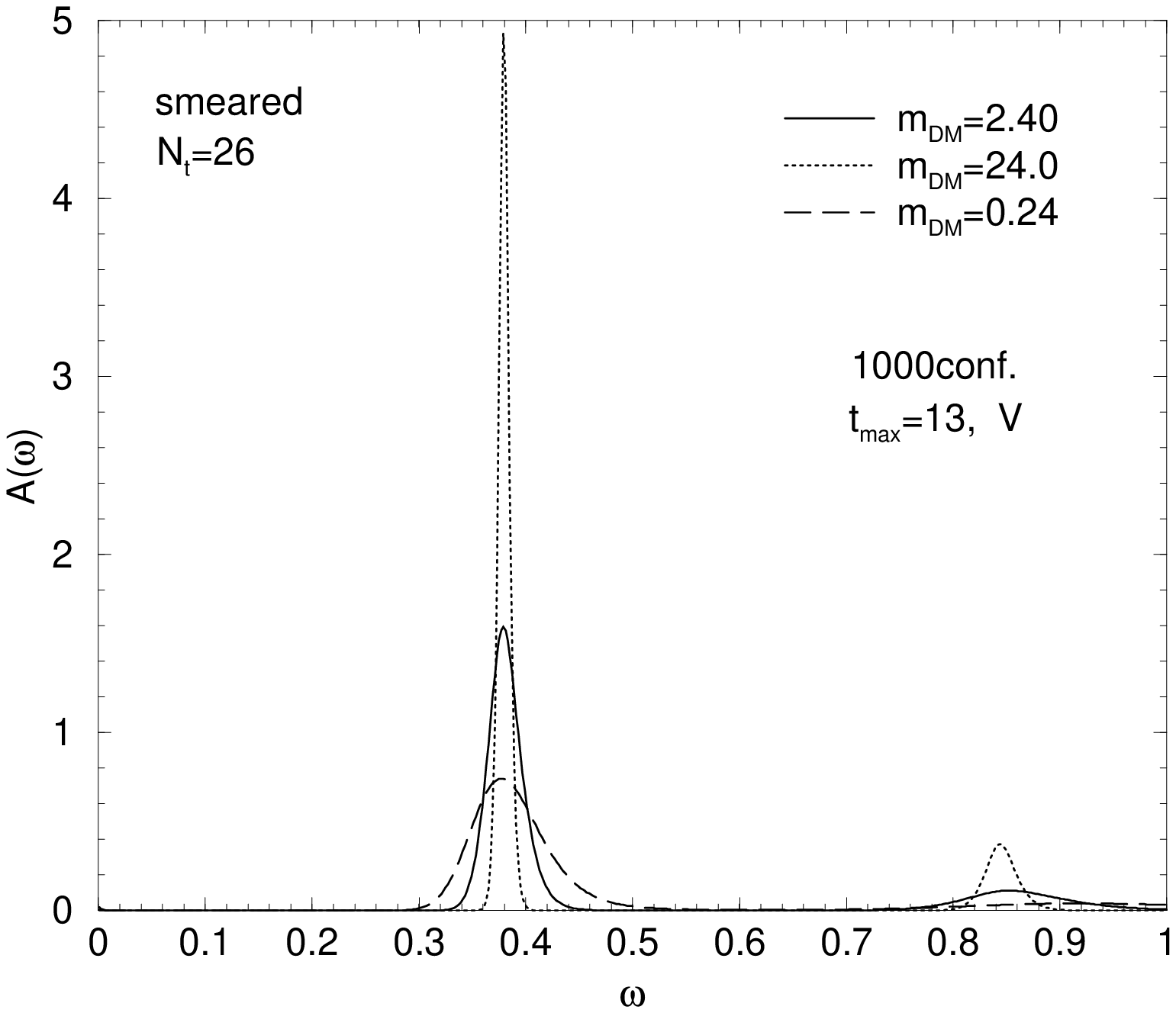}
\caption{
The spectral function at $N_t=26$ determined with MEM
(smeared correlators).}
\label{fig:mem_26}
\end{figure}

\begin{figure}
\vspace{0.1cm}
\includegraphics[width=8.0cm]{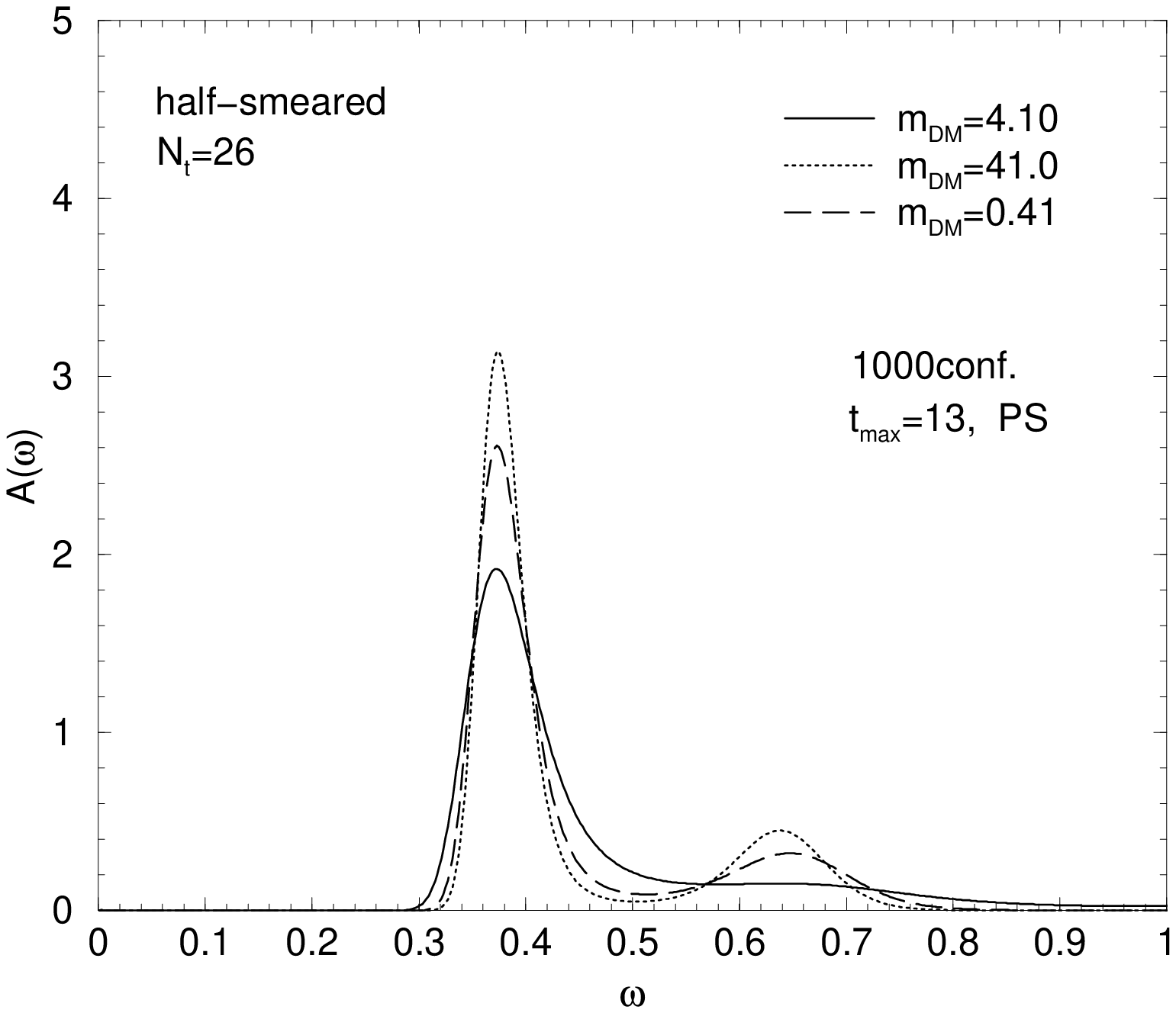}
\includegraphics[width=8.0cm]{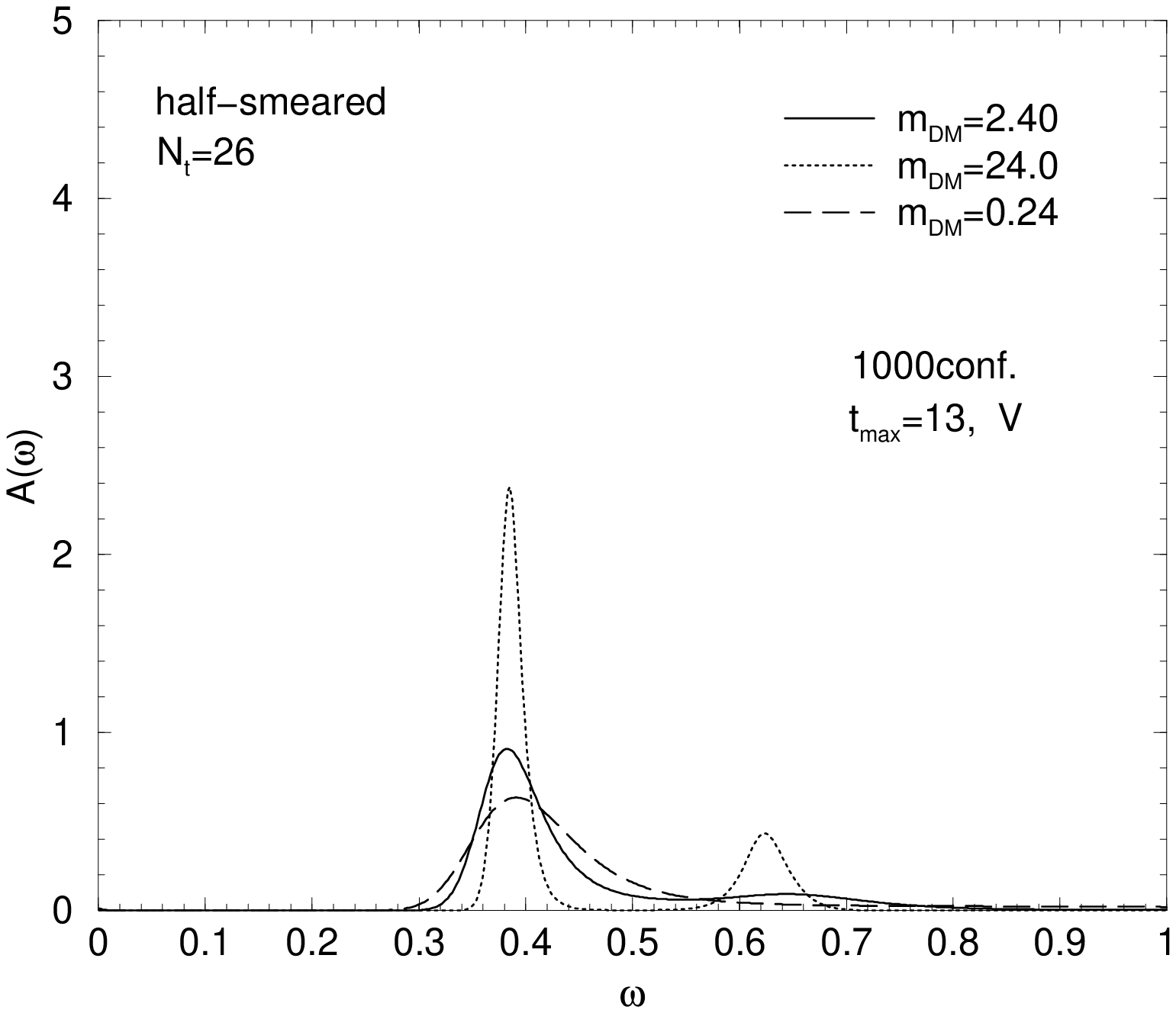}
\caption{
The results of MEM for the spectral function of the
half-smeared correlators at $N_t=26$.}
\label{fig:mem_26b}
\end{figure}

In this section, we analyze the correlators at
$T \simeq 1.1 T_c$ ($N_t=26$).

The setup of MEM analysis is the same as at $T\simeq 0.9 T_c$,
except for $t_{max}=13$.
Figure \ref{fig:mem_26} shows the result.
In both the PS and V channels there appear strong peak structures
around $\omega\simeq 0.4$.
This peak, hereafter called ``the ground state peak'' for
simplicity, appears at almost the same position as at $T\simeq 0.9$
but  with larger width.
Although, as a common tendency, the peak becomes sharper as the default
model parameter $m_{DM}$ increases, the position of the peak is unchanged.
It is also verified that the result is essentially the same with
less statistics, namely 500 configurations.
Therefore it is presumable that the spectral function still
has a peak structure at almost the same position as below $T_c$.

This result should be compared with the case where  the correlators
are composed of free quarks, considered in Sec.~\ref{subsec:free_quark}.
In the latter case, as shown in Sec.~\ref{subsec:analysis_free_quark},
MEM with the present number of degrees of freedom may produce the same
feature of the spectral function but with a lower peak position.
The above result therefore does not exclude the possibility that
the correlators are composed of two weakly interacting quarks
with rather large effective mass.

To judge between almost free quarks and genuine bound-state-like
structure we analyze the half-smeared correlator.
We repeat the same analysis as applied to the smeared correlators
for the half-smeared correlators and verify that MEM works up to
$t_{max}=12$ on almost the same level as for the smeared ones.
As discussed in Sec.~\ref{subsec:analysis_free_quark},
if the correlator is composed of two almost free quarks the
extracted spectral function should exhibit now a wider peak at a
higher  position.
Figure~\ref{fig:mem_26b} shows the result of MEM analysis
for the half-smeared correlators.
The analysis is performed in the same manner as for the smeared
correlators.
Comparing with the result for the smeared correlator, the
peak position for the half-smeared one is almost unshifted, while
the width of the peak tends to broaden slightly.
The latter effect , however, can also be easily explained as an effect of
default model uncertainty.
This result contradicts therefore the assumption that the peak
structure is an artifact of the smearing.
Hence the  MEM analysis supports a bound-state-like
structure at this temperature.

\subsection{Result of $\chi^2$ fit analysis}

\begin{figure}
\vspace{-0.4cm}
\includegraphics[width=9.0cm]{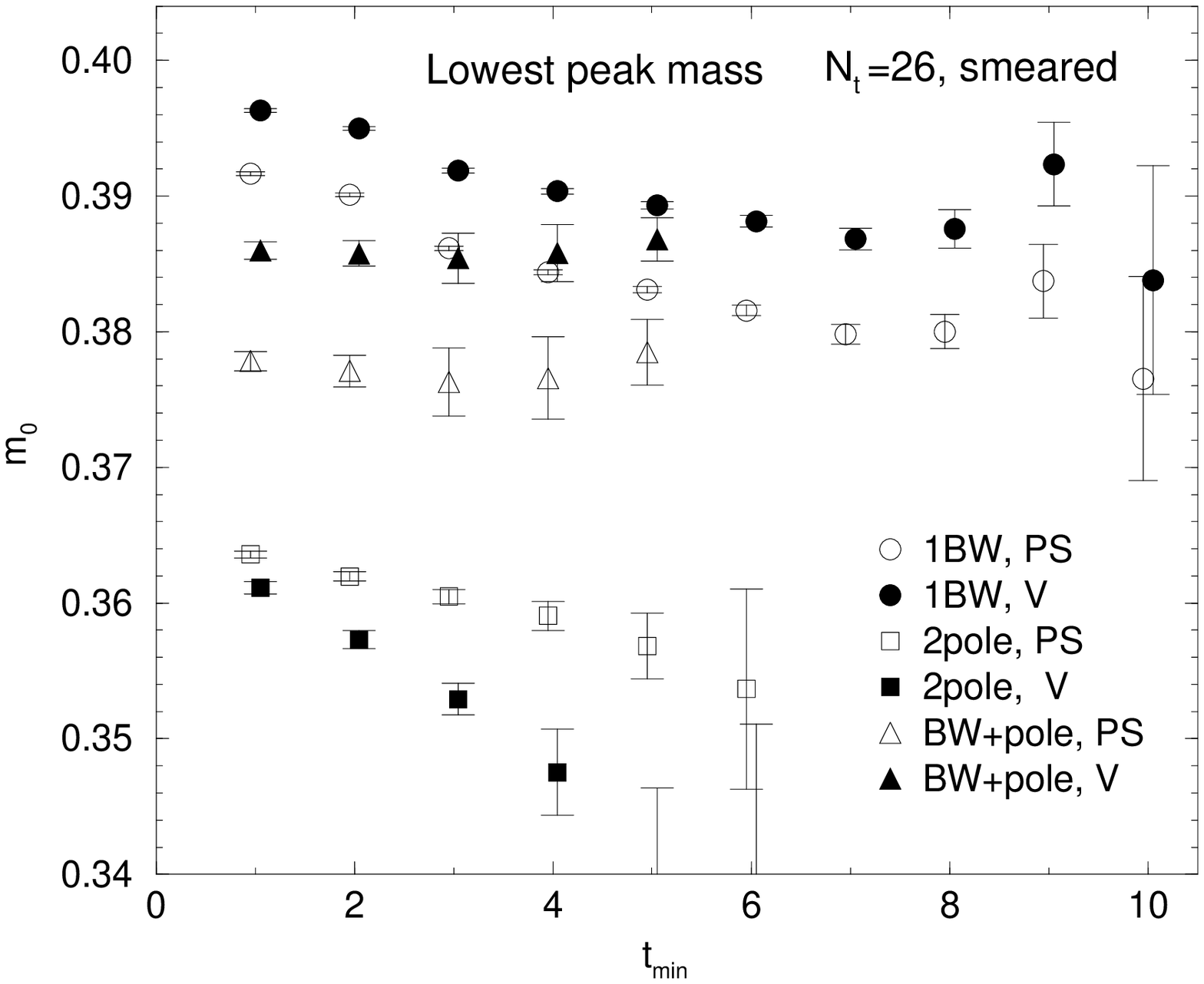}
\vspace{-1.2cm}\\
\includegraphics[width=9.0cm]{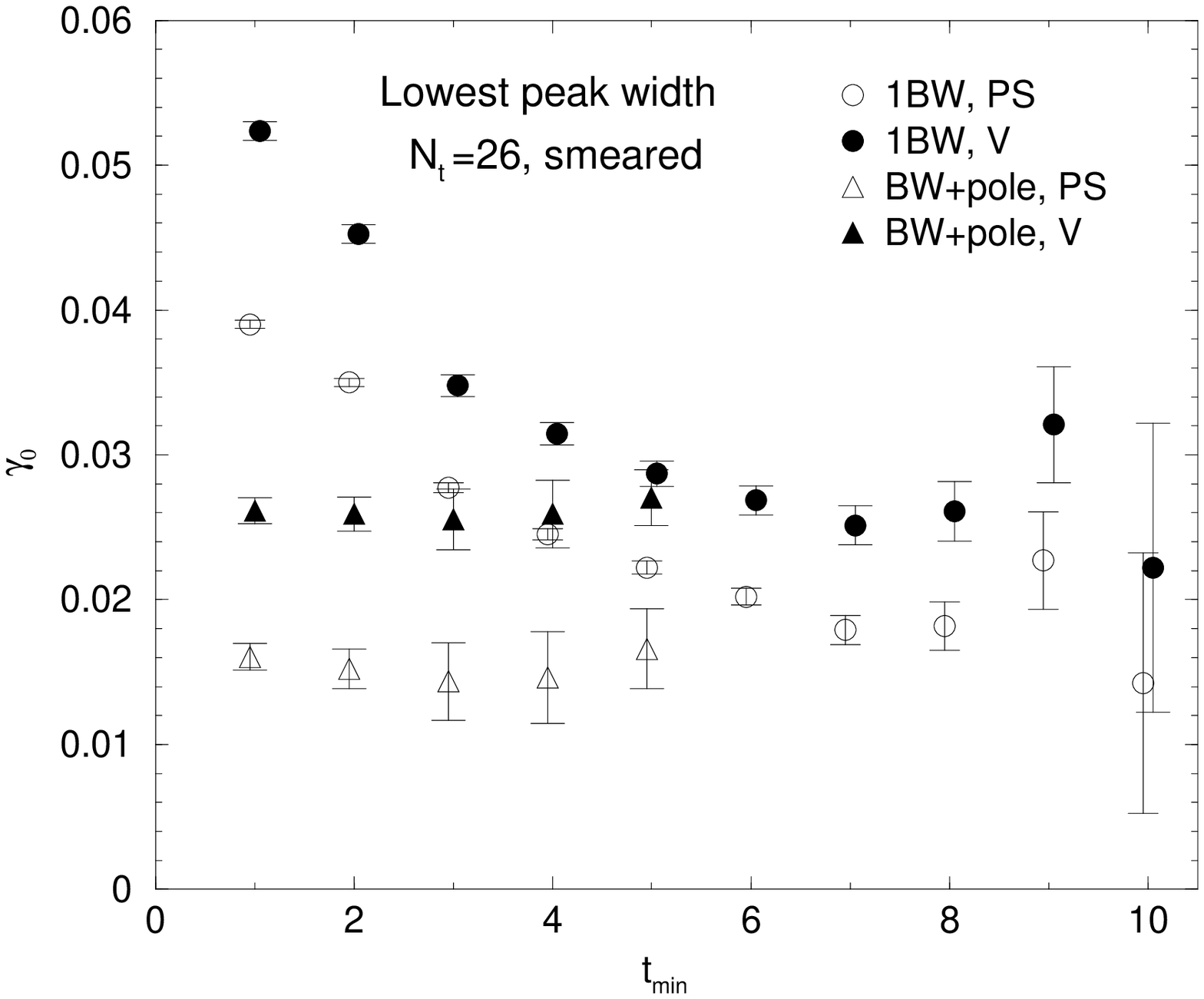}
\vspace{-0.8cm}
\caption{
The result of $\chi^2$ fit analysis for the smeared correlators
at $N_t=26$.}
\label{fig:fit_26}
\end{figure}

\begin{figure}
\vspace{-0.4cm}
\includegraphics[width=9.0cm]{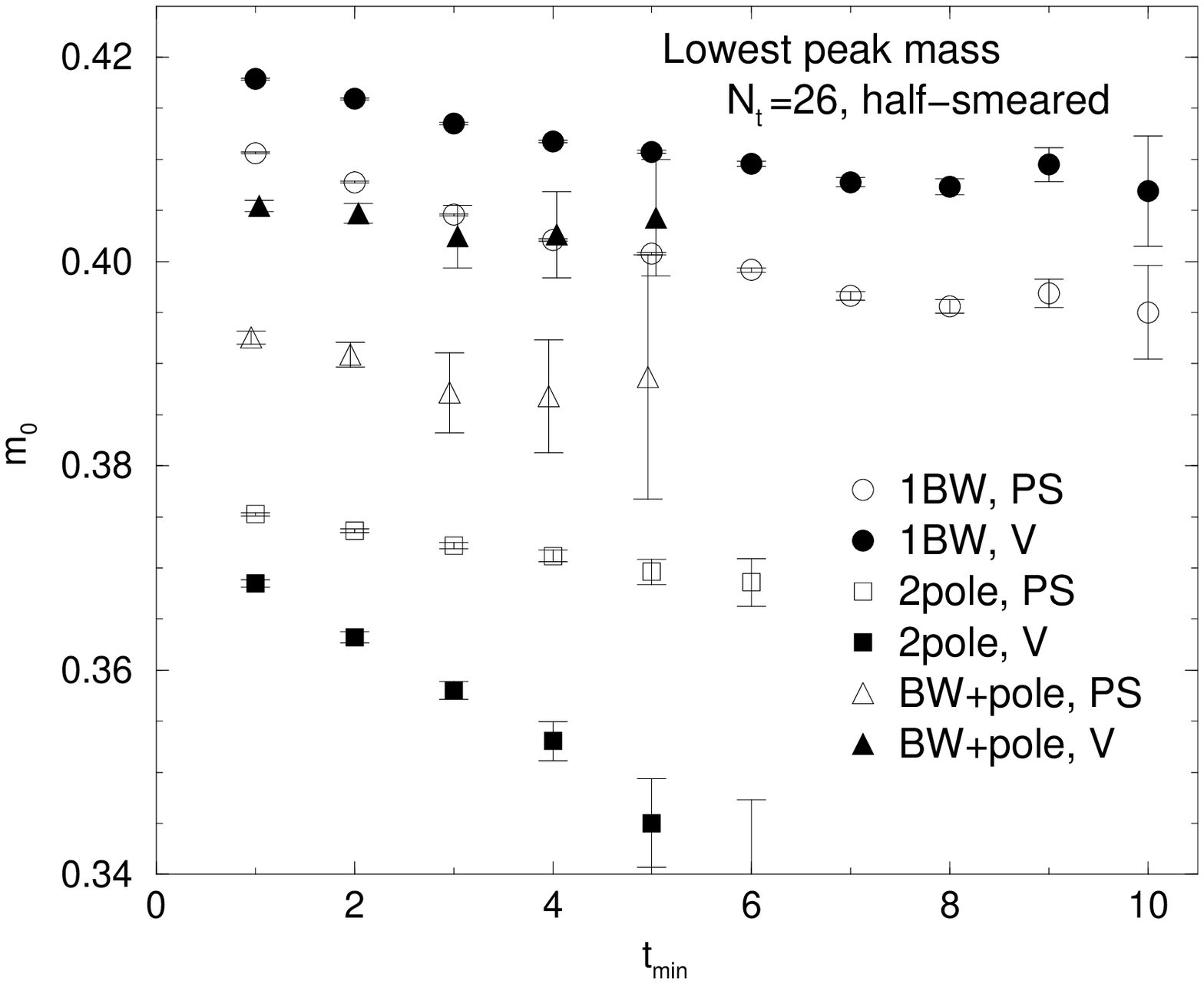}
\vspace{-1.2cm}\\
\includegraphics[width=9.0cm]{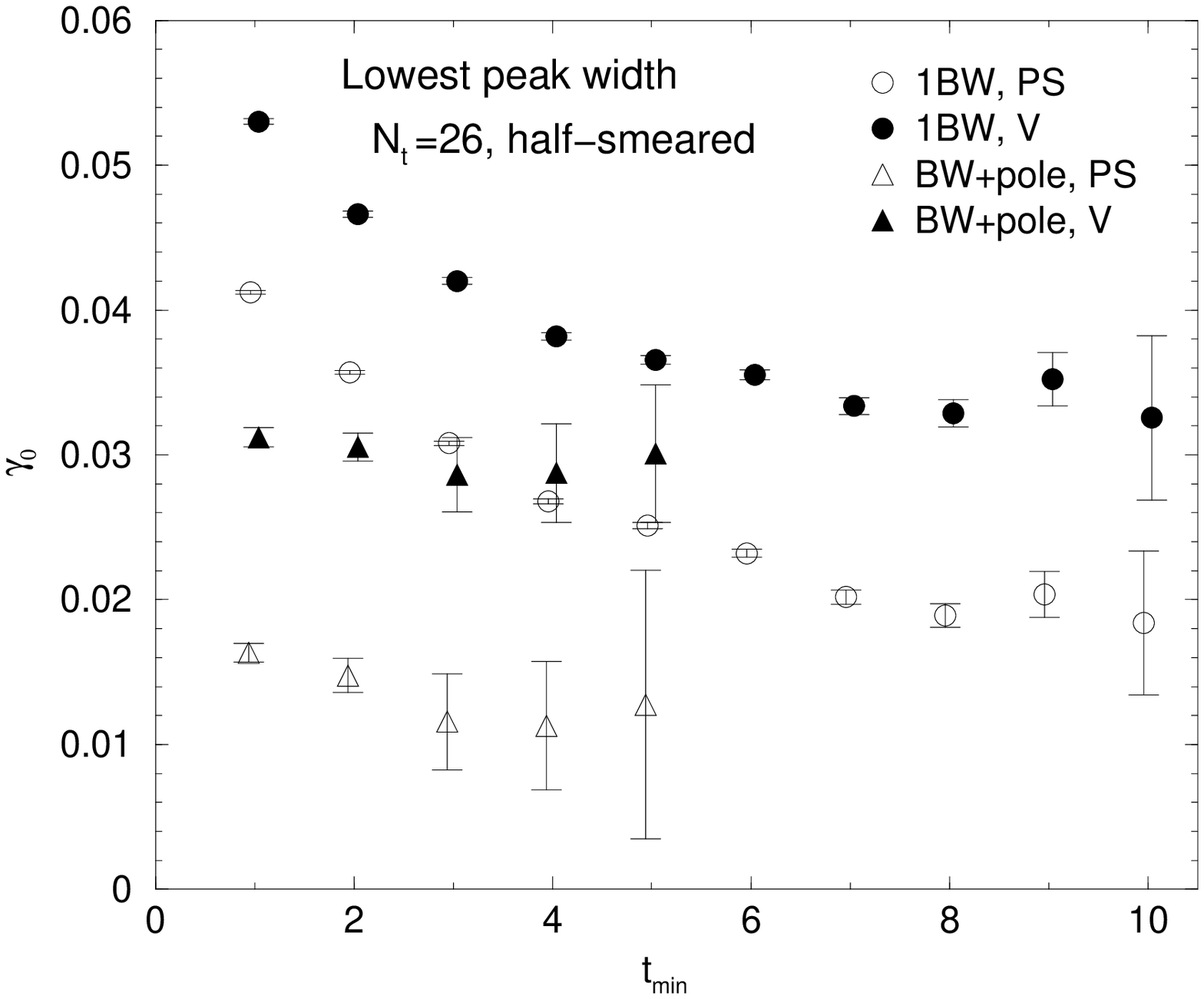}
\vspace{-0.8cm}
\caption{
The result of $\chi^2$ fit analysis for the half-smeared correlators
at $N_t=26$.}
\label{fig:fit_26b}
\end{figure}

\begin{figure}
\vspace{-0.35cm}
\includegraphics[width=9.2cm]{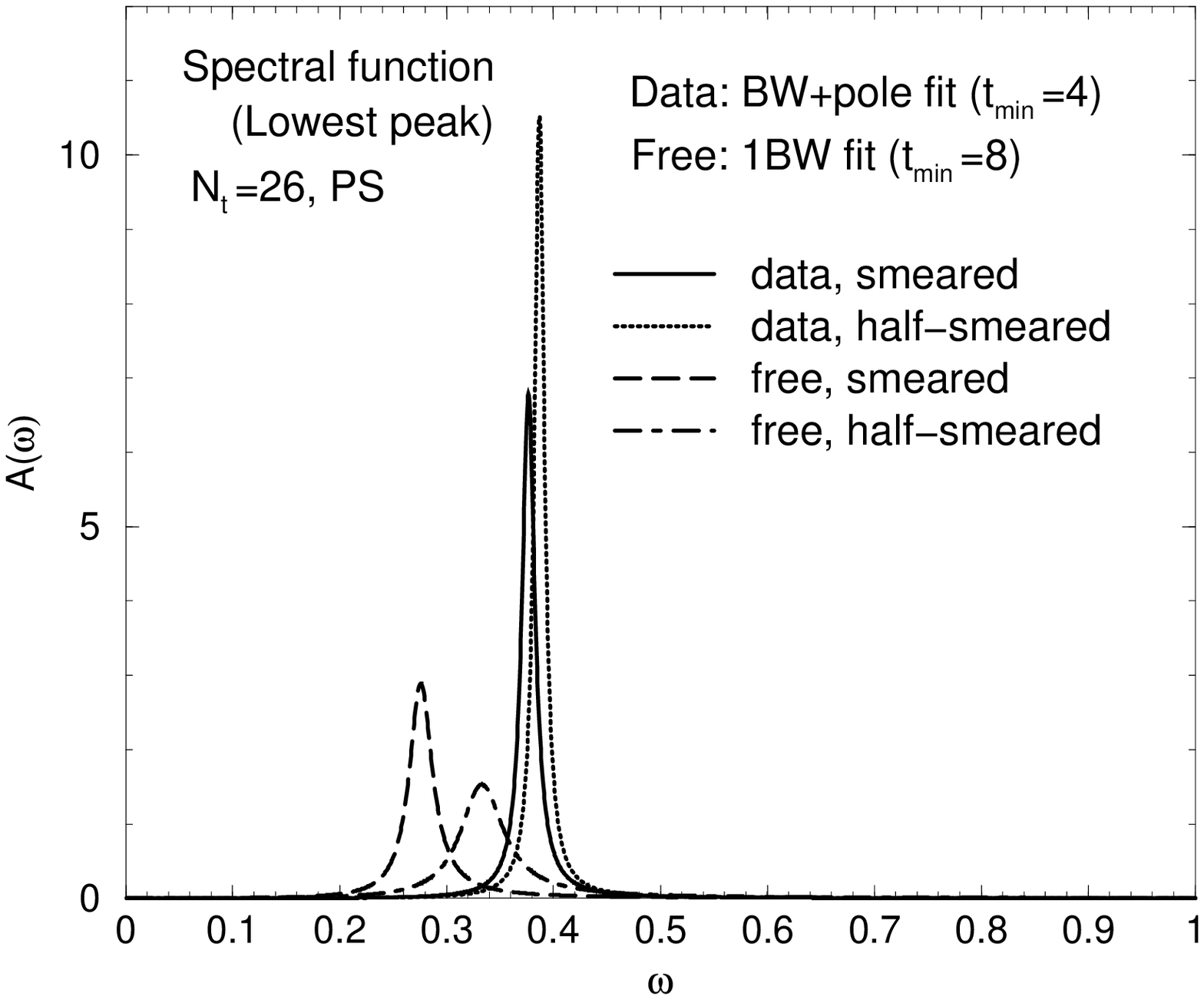}
\vspace{-1.2cm}\\
\includegraphics[width=9.2cm]{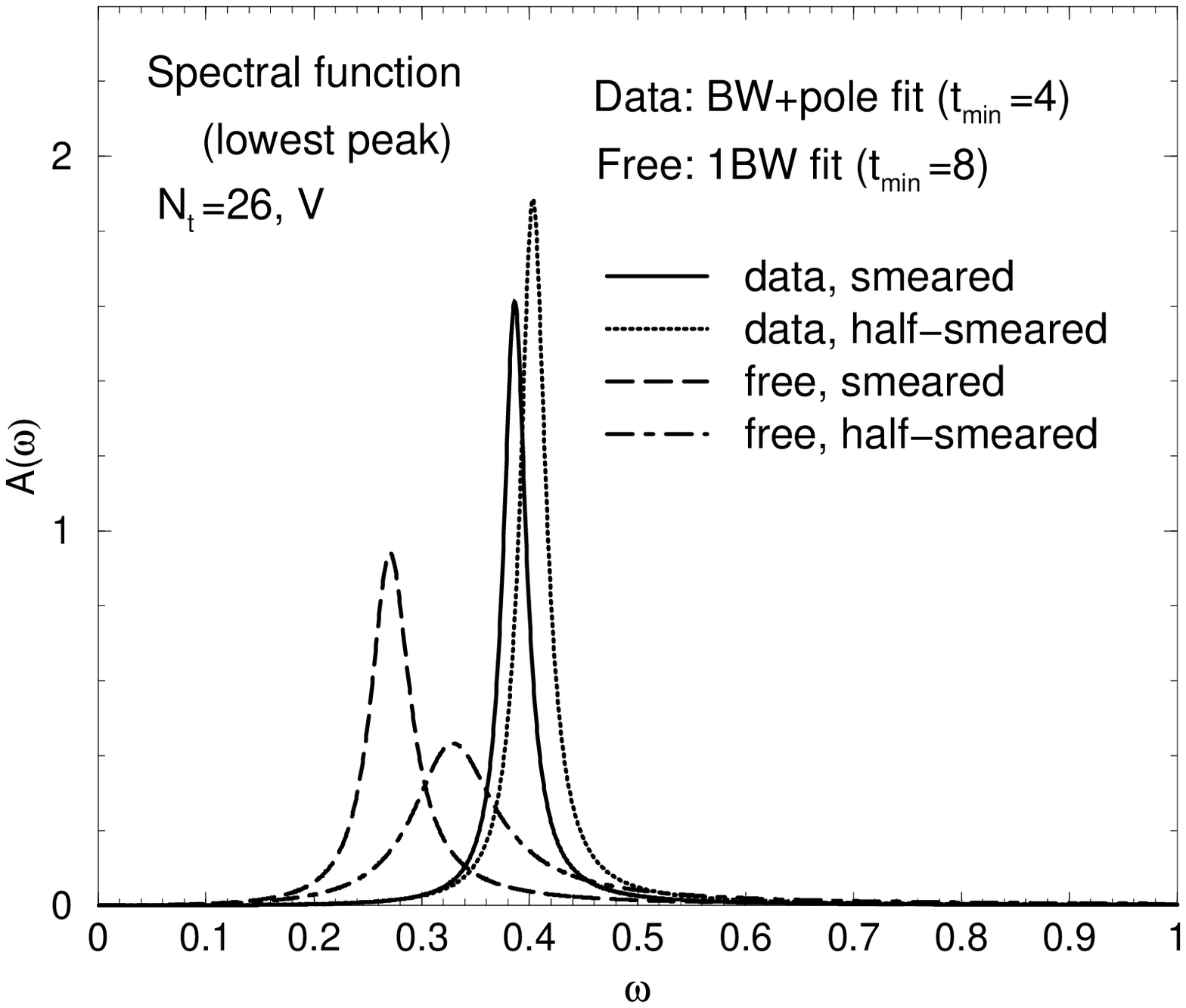}
\vspace{-0.9cm}
\caption{
The spectral functions obtained with $\chi^2$ fit analysis
at $N_t=26$.
For the correlators composed of free quarks only the result of 
BW fit is displayed, while a  BW+pole fit is used  for  the
correlators measured in the simulation.}
\label{fig:fit_26c}
\end{figure}

According to the result of MEM, it is reasonable to assume that
the spectral function at this temperature also have a peak structure,
similarly to the $T<T_c$ case.
Therefore the three forms for the $\chi^2$ fit analysis
are still reasonable assumptions to investigate the low energy
structures of the correlators.

Figure~\ref{fig:fit_26} shows the result of the $\chi^2$ fit analysis
for the smeared correlators.
The top panel shows the $t_{min}$ dependence of the mass parameters.
In contrast to the case below $T_c$, the masses from 2-pole fit 
no longer approach those of the other two fit forms
but fall monotonously.
The masses from 1-BW and BW+pole fits exhibit consistent behaviors,
indicating that these fits represent better the correlators. 
Actually the values of $\chi^2/N_{DF}$ are consistently fluctuating
around unity at $t_{min}>8$ for 1-BW fit, and in the whole range
of $t_{min}$ for BW+pole fit.
In the BW+pole fit, the mass parameters for the second peaks take values
around 0.9 and 1.0 for PS and V channels, respectively,
which is roughly consistent with the result of MEM.
The consistency of 1-BW and BW+pole fits also holds for the width
parameters, as displayed in the bottom panel of
Figure~\ref{fig:fit_26}.
These results indicate therefore that the widths associated with
the ground state peaks are finite for both the correlators in
PS and V channels.

We now analyze the half-smeared correlators
to examine whether this observation of peak structure
is an artifact of the smearing.
The analysis is performed in the same manner as for the (fully)
smeared correlators.
The result is displayed in Figure~\ref{fig:fit_26b}.
The mass parameters for the half-smeared correlators (top panel)
show a similar tendency as for the smeared correlators,
although the result of 1-BW and BW+pole fits approach slightly
larger values than those of smeared correlators.
The shifts of the masses are at most 5\% and can be explained
with the frequency dependence of the overlap $R(\omega)$ in
Eq.~(\ref{eq:spc_RBW}).
A similar tendency is also found for the width parameters
in the bottom panel of Fig.~\ref{fig:fit_26b}, which appear even more
consistent with those of smeared correlators  than
the mass parameters.
Therefore we find no significant shift of the peak position or
broadening of the peak width, such as expected for the case with almost
free quarks discussed in Sec.~\ref{subsec:free_quark}.

Figure~\ref{fig:fit_26c} illustratively compares these
spectral functions extracted with the $\chi^2$ fits
from four kinds of correlators: the smeared and half-smeared correlators,
numerically measured and composed of free quarks.
The results of the numerical simulation are of the BW+pole fit
with $t_{min}=4$.
For the free quark case the results of 1-BW fit with $t_{min}=8$
is displayed.
We note that the normalizations of the spectral functions are
not significant in the present analysis.
In both the PS and V channels the spectral functions for the
numerically observed correlators strongly peak  near $\omega=0.4$
with small widths.
The effect of the smearing seems not significant.
In contrast, the spectral functions for the correlators composed of
free quarks are strongly dependent on the applied smearing.
Both a sizable shift of mass and a broadening of width are observed
in this case.
It is difficult to exclude completely the possibility
that the correlators are composed of  almost free
quarks with rather large effective mass and nontrivial
dispersion relation.
However, it is quite suggestive that a nontrivial structure
indicating the existence of bound-state-like structure in the low
energy region of spectral function subsists at this value of $T$.

Summarizing the results of MEM and $\chi^2$ fit analysis
we conclude that the charmonium correlators in PS and V channels
at $T\simeq 1.1 T_c$ possess a  nontrivial peak structure 
in the low energy region.
As representative values of $m_0$ and $\gamma_0$,
we quote the results of the BW+pole fit analysis with $t_{min}=4$
for the smeared correlators:
\begin{eqnarray}
\mbox{PS}: & & m_0 = 3.06(2) \mbox{ GeV}, \ \ 
                \gamma_0 = 0.12(3) \mbox{ GeV}, \\
\mbox{V}:  & & m_0 = 3.13(2) \mbox{ GeV}, \ \
                \gamma_0 = 0.21(2) \mbox{ GeV}.
%
\end{eqnarray}
The quoted errors in $m_0$ and
$\gamma_0$ are only the statistical ones and do not include the 
error in $a_{\tau}^{-1}$.
As obvious from the discussions in this sections, these results should
contain systematic uncertainties of the order of 5\% due to
the analysis procedures, apart from other uncertainties such as
finite lattice artifacts and quenching effects.
Compared with the result at $T\simeq 0.9 T_c$, the widths at
$T\simeq 1.1 T_c$ is sizable, indicating a genuine temperature
effect in the deconfined phase.
On the other hand, the masses are almost unchanged.
This result is contradictory to the absence of bound states
argued  from the potential model approach
\cite{MS86,Potential_model}.

\section{Conclusion}
 \label{sec:conclusion}

The main goals of this paper were (1) to elucidate the technical
problems in extraction of the spectral function from lattice data
of $t$-correlators, and (2) to investigate the temperature effect
on the spectral function near the deconfining phase transition.
In the following, we summarize and discuss the results 
obtained in this paper.

As techniques to extract the information on the spectral function
we examined maximum entropy method (MEM) and the standard $\chi^2$
fit assuming suitable forms of the spectral function.
It is essential to check the reliability of the applied
methods to the systems in question.
Our condition for reliability at finite temperature
is that the methods reproduce the correct form of the spectral function
at $T=0$ when the $t$-interval is restricted to the one forced on us 
at $T>0$.
We examined MEM by applying it to the point and smeared correlators.
We find that MEM does not meet this requirement for the former, while
it does for the latter.
This is understandable, since the smearing enhances the
low energy part of the spectral function, which is what we are interested in,
while a much wide energy region contributes to the point correlators.
We note that whether MEM correctly works or not depends primarily 
on the extension of the physical $t$-region.
In particular, for the point correlators a region of $t$ of
$O(1 \mbox{fm})$ is necessary.
For the smeared correlator, this condition is much relaxed.
Therefore  only the smeared correlators were analyzed at finite
temperatures.

Since MEM is ambiguous concerning the quantitative detail
of the spectral function, the latter should be also analyzed with
other procedures.
As such a procedure, the $\chi^2$ fit is a reasonable candidate,
since MEM already gives a hint for a suitable {\it ansatz}
for the spectral function.
With several assumed forms and examining the fit range dependence
of the resultant values for the parameters,
this procedure gives us more quantitative information
on the properties of spectral function.
We emphasize that MEM and $\chi^2$ fit analysis as
used here are complementary
to each other.
This two-step approach actually worked for analyses of
the correlators at finite temperature, as well as at $T=0$.

Now we discuss the physical implications of our results
below and above the critical temperature.
We remind that our numerical simulation was performed without
dynamical quark effects.

Below the deconfining temperature, at $T\simeq 0.9 T_c$,
the reconstructed spectral function has a strong peak corresponding to
the ground state, with almost the same mass as at $T=0$ and
narrow width consistent with zero.
In contrast to the potential model analysis \cite{Has86},
the charmonium mass is not changed up to this temperature.
Similar tendencies have been reported in previous lattice QCD
calculations for the mesonic channels \cite{TARO01,Ume01},
while sizable reduction of mass has been observed for glueballs
\cite{ISM02}.
Considering the rather quantitative success of the potential model
approach for the charmonium systems at $T=0$ \cite{Cornel},
it is important to explain this discrepancy.

At $T \simeq 1.1 T_c$  we observed an indication that the spectral
functions still has strong peaks at almost the same positions
as $T<T_c$, and with widths of about 0.12 and 0.21 GeV for PS
and V channels, respectively.
This result presumably indicates the existence of quasi-stable
bound-state-like structures persistent up to this temperature.
The possibility of observing correlators composed of
almost free quarks (but of large effective masses) is, however, 
not completely excluded.
 Finally, also above $T_c$ the observations are not in accord with
the expectation from the potential model approach
\cite{Potential_model}.
This result implies that the plasma phase has a nontrivial
structure at least near the critical temperature.

For a more definite understanding of hadronic correlators at $T>0$
more studies containing dynamical simulations are necessary,
as well as calculations in wider range of temperatures.
The techniques adopted in this paper should be applicable to
these situations.
It is also important to repeat the same sort of analysis
in the light hadron sectors, for which a nontrivial structure
of correlators above $T_c$ was also reported \cite{TARO01}.

\section*{Acknowledgments}

We would like to express our sincere gratitude to Professor Osamu
Miyamura, who tragically passed away in July 2001, for
his inspiring us with this work and for valuable discussions.
We thank  N.~Ishii, T.~Kunihiro, A.~Nakamura, T.~Onogi, 
I.-O. Stamatescu, and T.~Yamazaki for useful discussions.
We are also grateful to the members of QCD-TARO Collaboration;
this work was done partly in accordance with their long
standing physical goals.
The simulation has been done on
NEC SX-5 at Research Center for Nuclear Physics, Osaka University and
Hitachi SR8000 at KEK (High Energy Accelerator Research Organization).
T.~U. is supported by the center-of-excellence (COE) program
at CCP, University of Tsukuba.
H.~M. is supported by Japan Society for the Promotion of Science
for Young Scientists.


\begin{thebibliography}{99}


\bibitem{Has86}
 T.~Hashimoto, O.~Miyamura, K.~Hirose and T.~Kanki, 
  Phys. Rev. Lett. {\bf 57}, 2123 (1986).

\bibitem{MS86}
 T.~Matsui and H.~Satz,
  Phys. Lett. B {\bf 178}, 416 (1986).

\bibitem{Jpsi:exp}
 NA50 Collaboration, M.~C.~Abreu {\it et al.},
  Phys. Lett. B {\bf 477}, 28 (2000).

\bibitem{Jpsi:th}
 For a review, H.~Satz,
   Nucl. Phys. B (Proc.\ Suppl.) {\bf 94}, 204 (2001).

\bibitem{AGDF59}
 A.A. Abrikosov, L.P.Gor'kov, and Dzyaloshinskii,
  Sov. Phys. JETP {\bf 36(9)}, 636 (1959);
 E.S. Fradkin,
  Sov. Phys. JETP {\bf 36(9)}, 912 (1959).

\bibitem{HNS93}
 T.~Hashimoto, A.~Nakamura and I.~O.~Stamatescu,
  Nucl. Phys. {\bf B400}, 267 (1993);
  Nucl. Phys. {\bf B406}, 325 (1993).

\bibitem{TARO01}
 QCD-TARO Collaboration, Ph. de Forcrand {\it et al.},
  Phys. Rev. D {\bf 63}, 054501 (2001).

\bibitem{Ume01}
 T.~Umeda, R.~Katayama, O.~Miyamura and H.~Matsufuru,
  Int. J. Mod. Phys. A {\bf 16}, 2215 (2001).

\bibitem{NAH99}
 Y.~Nakahara, M.~Asakawa and T.~Hatsuda,
  Phys. Rev. D {\bf 60}, 091503 (1999); 
 M.~Asakawa, T.~Hatsuda and Y.~Nakahara,
  Prog. Part. Nucl. Phys. {\bf 46}, 459 (2001). 

\bibitem{SpF_early}
 For pioneering works,
 QCD-TARO Collaboration, Ph. de Forcrand {\it et al.},
  Nucl. Phys. B (Proc. Suppl.) {\bf 63}, 460 (1998);
 E.~G.~Klepfish, C.~E.~Creffield, E.~R.~Pike,
  {\it ibid.}, 655 (1998).

\bibitem{WK00}  
 I.~Wetzorke and F.~Karsch,
  hep-lat/0008008.

\bibitem{ODS00}
 M.~Oevers, C.~Davies and J.~Shigemitsu,
  Nucl. Phys. B (Proc. Suppl.) {\bf 94}, 423 (2001).

\bibitem{CPPACS01a}  
 CP-PACS Collaboration, T.~Yamazaki {\it et al.},
  Phys. Rev. D {\bf 65}, 014501 (2002).

\bibitem{Fie02}
 H.~R.~Fiebig,
  Phys.\ Rev.\ D {\bf 65}, 094512 (2002);
 H.~R.~Fiebig, LHP collaboration,
  Nucl. Phys. B (Proc. Suppl.) {\bf 106}, 344 (2002).

\bibitem{AMR02}
 Difficulty in extracting transport coefficients from lattice data
  has been pointed out in:
  G.~Aarts and J.~M.~Martinez Resco,
   JHEP {\bf 0204}, 053 (2002);  
   hep-lat/0209033.

\bibitem{Bielefeld02a}  
 F.~Karsch, E.~Laermann, P.~Petreczky, S.~Stickan and I.~Wetzorke,
  Phys. Lett. {\bf B530}, 147 (2002);  
 I.~Wetzorke {\it et al.},
  Nucl. Phys. B (Proc. Suppl.) {\bf 106}, 510 (2002). 

\bibitem{Bielefeld02b}  
 P.~Petreczky, F.~Karsch, E.~Laermann, S.~Stickan and I.~Wetzorke,
  Nucl. Phys. B (Proc. Suppl.) {\bf 106}, 513 (2002).

\bibitem{Bielefeld02c}  
 S.~Datta, F.~Karsch, P.~Petreczky and I.~Wetzorke,
  hep-lat/0208012.

\bibitem{AHN02}   
 M.~Asakawa, T.~Hatsuda and Y.~Nakahara,
  hep-lat/0208059.

\bibitem{ISM02}
 N.~Ishii, H.~Suganuma and H.~Matsufuru,
 Phys. Rev. D {\bf 66}, 014507 (2002);
 hep-lat/0206020, to appear in Phys. Rev. D.

\bibitem{NMUM02}
 K.~Nomura, O.~Miyamura, T.~Umeda and H.~Matsufuru,
  hep-lat/0209139.

\bibitem{Kar82}
 F.~Karsch,
  Nucl. Phys. {\bf B205}, 285 (1982).

\bibitem{Aniso01a}
 J.~Harada, A.~S.~Kronfeld, H.~Matsufuru, N.~Nakajima and T.~Onogi,
  Phys. Rev. D {\bf 64}, 074501 (2001).

\bibitem{Aniso01b}
 H.~Matsufuru, T.~Onogi and T.~Umeda,
  Phys. Rev. D {\bf 64}, 114503 (2001).

\bibitem{LM93}
 G.~P.~Lepage and P.~B.~Mackenzie,
  Phys. Rev. D {\bf 48}, 2250 (1993).

\bibitem{KLM}  
 G.P.Lepage,
  Nucl. Phys. B (Proc. Suppl.) {\bf 26}, 45 (1992);
 P.~B.~Mackenzie,
  {\it ibid.} {\bf 30}, 35 (1993);
 A.~S.~Kronkeld,
  {\it ibid.} {\bf 30}, 445 (1993).

\bibitem{Kla98}
 T.~R.~Klassen,
  Nucl. Phys. {\bf B533}, 557 (1998).

\bibitem{Som94}
 R.~Sommer,
  Nucl. Phys. {\bf B411}, 839 (1994).

\bibitem{Aniso02a}
 J.~Harada, H.~Matsufuru, T.~Onogi and A.~Sugita,
  Phys. Rev. D {\bf 66}, 014509 (2002).

\bibitem{TARO02}
 QCD-TARO Collaboration, S.~Choe {\it et al.},
  Nucl. Phys. B (Proc.Suppl.) {\bf 106}, 361 (2002).

\bibitem{PDG00}
 Particle Data Group Collaboration, D.~E.~Groom {\it et al.},
  Eur. Phys. J. C {\bf 15} (2000) 1.

\bibitem{Potential_model}
 F.~Karsch, M.~T.~Mehr and H.~Satz,
  Z. Phys. C {\bf 37}, 617 (1988).

\bibitem{Cornel}
 E.~Eichten, K.~Gottfried, T.~Kinoshita, K.~D.~Lane and T.~M.~Yan,
  Phys. Rev. D {\bf 17}, 3090 (1978);
               {\bf 21}, 313(E) (1980);
               {\bf 21}, 203 (1980).

\end{thebibliography}
\end{document}